\pdfoutput=1
\documentclass[journal]{IEEEtran}
\usepackage{amsmath,amsfonts,amssymb}
\usepackage{mathrsfs}
\usepackage{mathtools}
\usepackage{enumitem}
\usepackage{multirow}

\usepackage[ruled,vlined]{algorithm2e}

\usepackage{array}
\usepackage[caption=false,font=normalsize,labelfont=sf,textfont=sf]{subfig}
\usepackage{textcomp}
\usepackage{stfloats}
\usepackage{url}
\usepackage{verbatim}
\usepackage{graphicx}
\usepackage{cite}

\usepackage[dvipsnames]{xcolor}
\usepackage{hyperref}
\hypersetup{
    colorlinks=true,
    citecolor=Green,
    linkcolor=blue,
    filecolor=magenta,
    urlcolor=cyan
}

\usepackage{siunitx}
\DeclareSIUnit\erg{erg}

\usepackage{notations}

\begin{document}
\bstctlcite{IEEEexample:BSTcontrol}

\title{
    Efficient sampling of non log-concave posterior\\
    distributions with mixture of noises
}

\author{
    \IEEEauthorblockN{
        Pierre Palud\IEEEauthorrefmark{1,2}, 
        Pierre-Antoine Thouvenin\IEEEauthorrefmark{1},
        Pierre Chainais\IEEEauthorrefmark{1},
        Emeric Bron\IEEEauthorrefmark{2},
        Franck Le Petit\IEEEauthorrefmark{2}\\
    }
    \vspace{2mm}
    \IEEEauthorblockA{\IEEEauthorrefmark{1}Univ. Lille, CNRS, Centrale Lille, UMR 9189 CRIStAL, F-59000 Lille, France} \\
    \IEEEauthorblockA{\IEEEauthorrefmark{2}LERMA, Observatoire de Paris, PSL Research University, CNRS, Sorbonne Universit\'es, 92190 Meudon, France} \\
}        




\maketitle

\begin{abstract}
This paper focuses on a challenging class of inverse problems that is often encountered in applications.
The forward model is a complex non-linear black-box, potentially non-injective, whose outputs cover multiple decades in amplitude.
Observations are supposed to be simultaneously damaged by additive and multiplicative noises and censorship.
As needed in many applications, the aim of this work is to provide uncertainty quantification on top of parameter estimates.
The resulting log-likelihood is intractable and potentially non-log-concave.
An adapted Bayesian approach is proposed to provide credibility intervals along with point estimates.
An MCMC algorithm is proposed to deal with the multimodal posterior distribution, even in a situation where there is no global Lipschitz constant (or it is very large).
It combines two kernels, namely an improved version of PMALA~\cite{li_preconditioned_2016} and a Multiple Try Metropolis (MTM) kernel~\cite{liu_multiple-try_2021}.
%
%
%
This sampler addresses all the challenges induced by the complex form of the likelihood.
The proposed method is illustrated on classical test multimodal distributions as well as on a challenging and realistic inverse problem in astronomy.
\end{abstract}


\section{Introduction}



%
Physics and experimental sciences often produce non-linear, potentially non-injective, forward models.
%
Such models are often encoded by expensive black-box functions, e.g., the solution of a large set of partial differential equations in epidemiology~\cite{walker_parameter_2010}
or astrophysics~\cite{wu_constraining_2018, joblin_structure_2018}.
The forward model may also span multiple decades, as in astrophysics~\cite{le_petit_model_2006} where orders of magnitude can be gigantic.
As a consequence, when the log-likelihood function is smooth, the Lipschitz constant of the gradient is too large to be numerically useful.
Inverse problems that involve such models can lead to a non-log-concave, potentially multimodal likelihood function.

For the sake of simplicity, most observation models consider one source of noise only,
%
%
while more detailed models may involve multiple noises as well as censored data, due to sensitivity limitations.
%
%
Such difficult models are often addressed by simplifying the scenario in practice.
For instance, one noise is assumed to dominate the others that are neglected, as in medical ultrasound imaging~\cite{krissian_oriented_2007} or in synthetic aperture radar~\cite{durand_multiplicative_2010}.


This work addresses a family of inverse problems involving both a non-linear black-box forward model covering multiple decades, and censored observations affected by both an additive and a multiplicative noise.
The log-posterior is intractable, admits a Lipschitz continuous gradient with a very large constant (if finite), may be non-concave and potentially multimodal.
A Bayesian approach is proposed. The problem is addressed with a Markov Chain Monte Carlo (MCMC) algorithm~\cite{robert_monte_2004,pereyra_survey_2016,luengo_survey_2020} to provide point estimates with the corresponding credibility intervals.
This uncertainty quantification is particularly critical for applications where no ground truth is available, as in cosmology and astrophysics~\cite{wu_constraining_2018, joblin_structure_2018}.
An explicit and smooth approximation of the likelihood is proposed.
It relies on a model reduction and on controlled approximations of the noise model. 
%
Since the Lipschitz constant of the gradient of the log-posterior is assumed to be very large or even infinite, efficient sampling methods relying on smoothness assumptions, such as the Metropolis Adjusted Langevin Algorithm~(MALA)~\cite{roberts_langevin_2002} or Hamiltonian Monte Carlo~(HMC)~\cite{brooks_mcmc_2011} typically fail to explore the parameter space.
To address this issue, a preconditioned MALA (PMALA) kernel~\cite{girolami_riemann_2011,xifara_langevin_2014} exploiting the {RMSProp preconditioner from deep-learning~\cite{rmsprop} is considered.
Compared to HMC and MALA, the exploration of the parameter space is based on local second order information of the log-posterior, instead of the global Lipschitz constant of its gradient only.
%
%
A first version of this kernel, introduced in~\cite{li_preconditioned_2016}, led to an approximate sampler. We further improve it in this paper to obtain an exact sampler, \emph{i.e.}, asymptotically drawing samples from the distribution of interest.
To account for the non-log-concavity and potential multimodality of the posterior, a combination of PMALA with a Multiple Try Metropolis (MTM) kernel~\cite{liu_multiple-try_2021} is proposed.

The proposed sampler is validated on two classical multimodal examples: a 2D Gaussian mixture model and the sensor localization problem~\cite{ihler_nonparametric_2005}.
It is then applied with good performances to a realistic higher dimensional astrophysical problem that combines all the aforementioned challenges. 
A preliminary version of this work was published in~\cite{palud_mixture_2022}, where the main principle of the method was summarized for a simplified illustration.

Section~\ref{section:model} introduces the Bayesian model, the proposed likelihood approximation and the resulting posterior distribution.
Section~\ref{section:mcmc} introduces the MCMC algorithm used to derive estimators.
Section~\ref{section:experiments} demonstrates the performance of the proposed method on the three experiments outlined above.
Section~\ref{section:conclu} provides conclusions and perspectives.

\section{Bayesian model}\label{section:model}
\IEEEpubidadjcol

This section introduces the general Bayesian model considered in this article.
A tractable surrogate likelihood with controlled error is built on a reduced forward model and a likelihood approximation that deals with the mixture of noises.
The prior and the resulting posterior are then introduced.

\subsection{Notation}

Throughout this paper, scalars are denoted with regular letters, e.g., indices $n$, $d$ and $\ell$, or the corresponding dimensions $N$, $D$ and $L$.
In the following, $N$ is the number of observations over $L$ channels and  $D$ is the number of parameters of the model.
Vectors are denoted using bold lowercase letters, e.g., parameters $\boldsymbol{\theta} \in \mathbb{R}^D$ or observations $\boldsymbol{y} = (y_1, \ldots, y_\ell, \ldots, y_L)\in \mathbb{R}^L$.
Matrices are written with bold uppercase letters, e.g., matrices of observations \mbox{$\boldsymbol{Y} = (\boldsymbol{y}_n)_{n=1}^N \in \mathbb{R}^{N \times L}$}.
The notation for functions is set accordingly, e.g., the forward model $\boldsymbol{f}(\boldsymbol{\theta}) = \left(f_\ell(\boldsymbol{\theta}) \right)_{\ell=1}^L$.

\subsection{Problem statement}

Individual observations $\obsvect{} = (y_\ell)_{\ell=1}^L$ gather $L$ channels.
They are considered to be generated from some parameter $\paramvect{} \in \R^D$ and a forward model $\truef : \R^D \to \R^L$, where the number $D$ of parameters is assumed to remain moderate, e.g. $D \lesssim 10$.
The forward model prediction for a channel $\ell$ is denoted by $\truefell$, so that for any $\paramvect{} \in \R^D$, $\truef(\paramvect{}) = \left(f_\ell(\boldsymbol{\theta}) \right)_{\ell=1}^L$.
The forward model $\truef$ is assumed to be a non-linear black-box function, e.g., the result of a physical experiment or a numerical simulation.
It is considered valid on a compact subset $\fvalidity = [l_1, u_1] \times \cdots \times [l_D, u_D] \subset \R^{D}$, with $l_d, u_d \in \R$ for any $d \in [\![ 1, D ]\!]$, that can correspond to a typical domain of validity.
To reflect physical considerations on the nature of the data, the function $\truef$ is further assumed to have positive values that can span multiple decades, as is often the case in astrophysics~\cite{le_petit_model_2006}.
%
%
Individual observations and parameters are grouped in indexed sets $\obsfull = (\obsvect{n})_{n=1}^N$ and $\paramfull = (\paramvect{n})_{n=1}^N$ of  size $N$, such as an image, a time series or more generally a graph, with $N$ potentially very large, e.g., of the order of millions.
The sensors are assumed to have a lower limit of sensitivity $\censor \in \R$ below which an observation is censored.
Both an additive and multiplicative noise degrade the observations.
%
Such a mixture of noises occurs in astrophysics as well as in medical ultrasound imaging~\cite{krissian_oriented_2007} or laser imaging and synthetic aperture radars~\cite{durand_multiplicative_2010} for instance. 
Turning to inference, one of the two noises is generally neglected for sake of tractability~\cite{krissian_oriented_2007}.
However, when the forward model spans several decades, the nature of the dominant noise depends on the amplitude of $\truefell(\paramvect{})$.
The resulting observation model is, for $n \in [\![1,N]\!]$ and $\ell \in [\![1,L]\!]$,
\begin{align} \label{eq:model}
    \obselt = \max \left\{
        \censor, \;
        \multnoise f_{\ell} (\paramvect{n}) + \addnoise
    \right\},
\end{align}
where $\addnoise \sim \mathcal{N}(0, \sigma_a^2)$ is an additive Gaussian white noise, and $\multnoise \sim \log \mathcal{N}(- \sigma_m^2 / 2, \sigma_m^2)$ is a lognormal multiplicative noise such that $\mathbb{E}[\multnoise] = 1$.
The noise terms $\addnoise$ and $\multnoise$ are assumed independent with known variances $\sigma_a^2$ and $\sigma_m^2$, respectively.
They are also assumed independent of $\truefell(\paramvect{n})$.
At low intensities, the additive noise dominates; high intensities are mainly damaged by the multiplicative noise.

%
%


\subsection{Likelihood approximation}
\label{subsection:likelihood}

The likelihood associated to~\eqref{eq:model} involves a potentially expensive forward model.
A model reduction can be used to ensure the computational efficiency of the inference process.
The presence of the two sources of noise makes the likelihood intractable so that we propose a parametric surrogate model.

\subsubsection{Model reduction}
\label{subsub:model_reduction}

The forward model $\truef$ is assumed to be encoded by an expensive black-box function.
%
Such black-box models may be addressed with a likelihood-free method such as Approximate Bayesian Computation (ABC)~\cite{beaumont_approximate_2002} that yield approximate samplers of the true posterior distribution.
These methods are limited by the cost of numerous evaluations of the black-box function.
A cheaper reduced model is preferred when this cost becomes prohibitive~\cite{peterson_zonal_2017,kwan_cosmic_2015}.
This solution often permits to exactly sample from an approximate posterior.
%
%
%
%
Model reduction largely remains an application specific problem, with only a few generic approaches~\cite{kasim_building_2021,bobin_non-linear_2021}.
Here the forward model $\truefell$ is positive and covers several decades, for all $\ell \in [\![1, L]\!]$. We propose to replace it by approximations $\approxPell$ of its logarithm so that
\begin{equation}\label{eq:approxfell}
    \forall \ell, \quad
    \approxfell(\paramvect{})
    = \exp \left[
        \approxPell(\paramvect{})
    \right]
    .
\end{equation}
%
The error introduced by replacing $\truef$ by $\approxf{}$ should be negligible compared to $\truef$ and to the noise standard deviations $\sigma_a$ and $\sigma_m$.
In the present approach, evaluations of $\approxPell$ and its gradients should be fast, as each iteration of the proposed MCMC algorithm in Section~\ref{section:mcmc} will require such operations.
This approximation is also required to be twice differentiable to satisfy the requirements of the PMALA kernel involved in~\ref{sec:PMALA}.
The derivation of such a reduced model is feasible with a wide family of methods including polynomials or neural networks but remain out of the scope of this work.
In the following, $\approxf$ is assumed available so that from now on $\truefell$ is replaced by $\approxfell$.

\subsubsection{Modeling the noise mixture}
\label{ssub:noise_approx}

For simplicity, we first consider the uncensored part of the model~\eqref{eq:model}.
Since the corresponding likelihood is intractable, most approaches in the literature~\cite{krissian_oriented_2007,durand_multiplicative_2010} neglect one source of noise.
This strategy obviates the need for handling both the additive and multiplicative noises at intermediate intensities.
A slightly different mixture model is addressed in~\cite{huang_convex_2013} with a hierarchical approach and linear forward models.
The proposed approach builds on~\cite{nicholson_additive_2020}, where the mixture is approximated with a purely additive model.
The additive noise $\addnoise$ in \eqref{eq:model} can be neglected when $\approxfell(\paramvect{n}) \to \infty$, while the multiplicative noise $\multnoise$ becomes negligible as $\approxfell(\paramvect{n}) \to 0$.
Therefore, for each observation $\obselt$, the true likelihood is approximated using three different regimes: low, intermediate and high values of $\approxfell(\paramvect{n})$.
In the low value regime, the true likelihood function $\pi(\obselt \vert \paramvect{n})$ is approximated by an additive Gaussian approximation $\pi^{(a)}(\obselt \vert \paramvect{n})$ corresponding to
\begin{align}
    \obselt \simeq \approxfell(\paramvect{n}) + e^{(a)}_{n,\ell},
    \quad
    e^{(a)}_{n,\ell} \sim \mathcal{N}(m_{a,n,\ell}, s_{a,n,\ell}^2)
    ,
\end{align}
where $m_{a,n,\ell}$ and $s_{a,n,\ell}^2$ are obtained by matching the two first moments with model~\eqref{eq:model}, which yields
\begin{align}\label{eq:def_m_a_s_a}
    \begin{cases}
      m_{a,\ell,n} = 0, \\
      s_{a,\ell,n}^2 =
      \approxfell(\paramvect{n})^2 (e^{\sigma_m^2} - 1) + \sigma_a^2
      .
    \end{cases}
\end{align}
Conversely, in the high value regime, a multiplicative lognormal approximation $\pi^{(m)}(\obselt \vert \paramvect{n})$ is used.
It reads
\begin{align}
    \obselt \simeq e^{(m)}_{n,\ell} \approxfell(\paramvect{n}),
    \quad
    e^{(m)}_{n,\ell} \sim \log \mathcal{N}(m_{m,n,\ell}, s_{m,n,\ell}^2)
    ,
\end{align}
where moment matching with~\eqref{eq:model} yields: 
\begin{align}\label{eq:def_m_m_s_m}
    \begin{cases}
        m_{m,\ell,n} =
        - \frac{1}{2} \left\{
            \sigma_m^2 +
            \log \left[
                1
                + \frac{\sigma_a^2}{\approxfell(\paramvect{n})^2 e^{\sigma_m^2}}
            \right]
        \right\}
        , \\
        s_{m,\ell,n}^2 =
        - 2 \, m_{m,\ell,n} \quad \text{so that } \mathbb{E}[e^{(m)}_{n,\ell}] = 1
        .
    \end{cases}
\end{align}

For the intermediate regime, for each channel $\ell$, we introduce parameters $\aell = (\lambdaLowerLimit, \lambdaUpperLimit) \in \R^2$.
$\lambdaLowerLimit$ pinpoints the low to intermediate value transition and $\lambdaUpperLimit$ the intermediate to high value transition.
In this intermediate regime, i.e., $\lambdaLowerLimit \leq  \approxPell(\paramvect{n}) \leq \lambdaUpperLimit$, we propose to use a geometric average of the two likelihood approximations $\pi^{(a)}(\obselt \vert \paramvect{n})$ and $\pi^{(m)}(\obselt \vert \paramvect{n})$ with weights $1 - \lambda$ and $\lambda$, respectively, see the first term of \eqref{eq:likelihood} below.
The weight function $\lambda$ is defined as a twice differentiable sigmoid with values in $[0, 1]$:
%
%
\begin{equation}
\label{eq:def_lambda}
    \lambda (\paramvect{n}, \aell) 
    = 
    \begin{cases}
        0 & \text{if } \approxPell (\paramvect{n}) \leq \lambdaLowerLimit \\
        1 & \text{if } \approxPell (\paramvect{n})  \geq \lambdaUpperLimit \\
        Q \left( \frac{\approxPell (\paramvect{n})  - \log \lambdaLowerLimit }{\log \lambdaUpperLimit - \log \lambdaLowerLimit}\right) &
        \text{otherwise}
    \end{cases},
\end{equation}
where $Q$ is a polynomial such that $Q(0) = 0$, $Q(0) = 1$ and $Q^{'}(0) = Q^{'}(1) = Q^{''}(0) = Q^{''}(1) = 0$ for $ \lambda$ to be $\mathscr{C}^2$.
One of the simplest such polynomials is $Q(u) = u^3 (6 u^2 - 15 u + 10)$.
%

To take censorship into account, let $\mathbf{C} = (c_{n,\ell})_{n, \ell} \in \{0, 1 \}^{N L}$ be a matrix such that $c_{n,\ell} = 1$ for a censored observation, and $c_{n,\ell} = 0$ otherwise.
Let $F^{(a)}(\cdot \vert \paramvect{n})$ and $F^{(m)}(\cdot \vert \paramvect{n})$ be the cumulative density functions (cdf) of $\pi^{(a)}(\cdot \vert \paramvect{n})$ and $\pi^{(m)}(\cdot \vert \paramvect{n})$, respectively.
The likelihood of censored data involves $F^{(a)}(\censor \vert \paramvect{n})$ and $F^{(m)}(\censor \vert \paramvect{n})$.
The proposed likelihood approximation of model~\eqref{eq:model} finally reads
\begin{align}\label{eq:likelihood}
    \tilde{\pi}
    & (\obselt \vert \paramvect{n},\aell)
    \propto \\
    &
    \left[ \pi^{(a)}(\obselt \vert \paramvect{n})^{1-\lambda(\paramvect{n}, \aell)} \;
    \pi^{(m)}(\obselt \vert \paramvect{n})^{\lambda(\paramvect{n}, \aell)} \right]^{1 - c_{n,\ell}} \nonumber \\
    & \times \left[ F^{(a)}(\censor \vert \paramvect{n})^{1-\lambda(\paramvect{n}, \aell)} \;
    F^{(m)}(\censor \vert \paramvect{n})^{\lambda(\paramvect{n}, \aell)} \right]^{c_{n,\ell}}  \nonumber
    .
\end{align}

The accuracy of this likelihood approximation clearly depends on the choice of the parameter $\aell$.
Appendix~\ref{section:llh_param_optim} proposes a procedure to adjust it in a relevant manner.

\subsection{Prior and resulting posterior}

We will consider applications on multispectral images so that this work combines two penalties to build the prior distribution.
The first one favors the spatial regularity of estimations.
It is based on a local regularizer $h : \R^N \to \R_+$ applied to each map $\paramfull_{\cdot d} = (\theta_{n,d})_{1 \leq n \leq N}$, with $d \in [\![ 1, D]\!]$.
The regularizer can be
the Euclidean norm of the usual gradient or Laplacian of the component map, with regularization parameter $\tau_d > 0$.
The second term encodes the validity of $\truef$ on a compact set $\fvalidity = [l_1, u_1] \times \cdots \times [l_D, u_D]$.
Note that the reduced model may be defined out of $\fvalidity$ but will not be considered as valid since it was not trained on such points.
The most natural approach would be to use the indicator function $\iota_{\fvalidity^N}$ of the set $\fvalidity^N$,
where $\iota_{\fvalidity^N}(\paramfull) = 0$ if $\paramfull \in \fvalidity^N$, and $+\infty$ otherwise.
Since the PMALA kernel to be introduced in Section~\ref{sec:PMALA} requires twice differentiability, the following twice differentiable approximation of $\iota_{\mathcal{C}^N}$, known as the quartic penalty in constrained optimization~\cite{nocedal_numerical_2006}, is used:
%
\begin{align}\label{eq:_neg_log_smooth_indicator_prior}
    \tilde{\iota}_{\fvalidity^N} :
    \paramfull 
    \mapsto
    \sum_{n=1}^N \sum_{d=1}^D
    \left[
        \max (0, \paramelt{n,d} - u_d, l_d - \paramelt{n,d})
    \right]^4
    ,
\end{align}
%
%
%
leading to a smooth uniform distribution.
Finally, the resulting prior distribution is given by
\begin{align}
    \label{eq:overall_prior}
    \pi(\paramfull) \propto 
    \exp \left(
        - \delta \; \tilde{\iota}_{\fvalidity^N} (\paramfull)
        - \sum_{d=1}^D \tau_d \; h(\paramfull_{\cdot d})
    \right)
    ,
\end{align}
%
where $\delta > 0$ is a penalty parameter.
%
%
The posterior distribution combines $N L$ independent likelihoods~\eqref{eq:likelihood} and the priors~\eqref{eq:overall_prior}.
%
%
\begin{align}
    \pi (\paramfull \vert \obsfull)
    & \propto \exp \left[ - g(\paramfull) \right] \label{eq:neglog_posterior} \\
    & \propto
    \left[
        \prod_{n=1}^N \prod_{\ell=1}^L
        \tilde{\pi}
        (\obselt \vert \paramvect{n}, \aell)
    \right]
    \pi(\paramfull) \label{eq:posterior}
    .
\end{align}
%
%
%

\section{Proposed MCMC sampler}\label{section:mcmc}

MCMC algorithms can provide point estimates along with the associated uncertainty quantification.
%
However, the posterior distribution of complex systems is in general non-log-concave, hence potentially multimodal, which makes the sampling task challenging.
In addition, when the forward model spans several decades, the gradient of the negative log-posterior $\nabla g$ has a potentially very large Lipschitz constant, if any.
To address these two challenges, a new transition kernel is proposed as a combination of two kernels: PMALA~\cite{xifara_langevin_2014} and MTM~\cite{liu_multiple-try_2021}.
PMALA tackles the regularity issue to efficiently explore the neighborhood of a local mode,
whereas MTM permits jumps between modes.

\subsection{PMALA transition kernel}
\label{sec:PMALA}

In absence of an exploitable gradient-Lipschitz regularity of the log-posterior $g$ in \eqref{eq:neglog_posterior}, a preconditioned MALA equipped with RMSProp~\cite{rmsprop} is introduced to perform an efficient local exploration of the posterior distribution.
To simplify notations, we temporarily use the vector version of $\paramfull$ in lexicographic order so that $\paramfull \in \R^{ND}$.
Metropolis-Hastings~(MH)~\cite{metropolis_equation_1953,hastings_monte_1970} is arguably the most famous MCMC algorithm.
At each step $t$, a candidate $\paramfull_c^{(t)}$ is sampled from a proposal distribution $q(\paramfull_c^{(t)} \vert \paramfull^{(t-1)})$ that is accepted with probability
\begin{align}\label{eq:accept_proba}
    \rho^{(t)} = 1 \wedge
    \frac{\pi \left(\paramfull_c^{(t)} \vert \obsfull \right)}{\pi \left(\paramfull^{(t-1)} \vert \obsfull \right)}
    \frac{q \left(\paramfull^{(t-1)} \vert \paramfull_c^{(t)} \right)}{q \left(\paramfull_c^{(t)} \vert \paramfull^{(t-1)} \right)}
    .
\end{align}
%
The random walk proposal is often used but does not scale up well due to its blind nature~\cite{pereyra_survey_2016}.
Hamiltonian Monte Carlo~(HMC)~\cite{brooks_mcmc_2011} and Metropolis Adjusted Langevin Algorithm~(MALA)~\cite{roberts_langevin_2002} both exploit gradient information.
MALA is defined as a discretized Langevin diffusion process with an accept-reject step, while HMC relies on Hamiltonian dynamics and auxiliary variables.
Both propose larger steps than the random walk with high acceptance probability, which favors scaling up~\cite{pereyra_survey_2016}.
They both rely on a step size inversely proportional to the Lipschitz constant of $\nabla g$, if it exists.
Here the forward model $\approxf{}$ covers several decades so that this Lipschitz constant is potentially very large or even infinite.
Therefore, MALA and HMC will typically fail to efficiently explore the parameter space of a posterior distribution such as~\eqref{eq:posterior}.

A transition kernel that handles such situations relies on extensions of HMC and MALA to Riemannian manifolds~\cite{girolami_riemann_2011}.
The MALA extension is favored over its HMC counterpart as MALA yields faster individual iterations and requires less parameter tuning.
Moreover, Riemannian manifolds MALA was improved in~\cite{xifara_langevin_2014}, resulting in the so-called position dependent MALA (PMALA) kernel.
It permits exploiting local information geometry thanks to a position dependent preconditioner.
%
%
%
We propose to use the RMSProp preconditioner~\cite{rmsprop} that was initially defined in the deep learning literature for fast neural networks training.
In the absence of exploitable Lipschitz constant, it adaptively estimates a local variance of the gradient $\nabla g$ by keeping memory of former proposals $\paramfull^{(t)}_c$.
At each step $t$, it updates a surrogate gradient variance vector $\velocity^{(t)} \in \R^{ND}$ such that for all $i \in [\![1, ND]\!]$,
\begin{align}
    \label{eq:used_v_pmala}
    v_i^{(t)}
    & = \alpha v_i^{(t - 1)} + (1 - \alpha) \left[
         \frac{\partial \neglogpost}{\partial \paramelt{i}} \left( \paramfull_c^{(t)} \right)
    \right]^2 \\
    & = (1-\alpha) \sum_{j=1}^t \alpha^{t-j} \left[
        \frac{\partial \neglogpost}{\partial \paramelt{i}} \left( \paramfull_c^{(t-j)} \right)
    \right]^2,
\end{align}
where $\alpha \in ]0, 1[$ is an exponential decay rate.
%
Note that the variance vector $\velocity^{(t)}$ relies on candidates $\paramfull_c^{(t)}$ instead of iterates $\paramfull^{(t)}$:
candidates might not be kept in the Markov chain, but they still contain important information about the shape of the distribution.
%
%
The RMSProp preconditioner is defined as~\cite{rmsprop}
\begin{equation}\label{eq:preconditioner}
    \preconditioner^{(t)}
    = \text{diag} \left(
        \frac{1}{\eta + \sqrt{\velocity^{(t)}}}
    \right)
    \in \R^{ND \times ND},
\end{equation}
with $\eta$ a small damping parameter.
This preconditioner has already been used in a MCMC context~\cite{li_preconditioned_2016} within an approximate sampler.
The goal in~\cite{li_preconditioned_2016} was to sample from a distribution defined over the parameters of a neural network trained on a large dataset.
Accept or reject steps were omitted as they would have required expensive computations on the full dataset.
Additionally, the discretization of the Langevin diffusion process equipped with a position-dependent preconditioner comes with an additional drift term~\cite{xifara_langevin_2014} that was neglected in~\cite{li_preconditioned_2016}.
We correct these two approximations to sample exactly from~\eqref{eq:posterior}.
Following~\cite{xifara_langevin_2014}, the proposal distribution corresponding to PMALA with the RMSProp preconditioner is the Gaussian distribution:
\begin{align}
    q \left( \paramfull_c^{(t)} \vert \paramfull^{(t-1)} \right) = \mathcal{N} \left( \paramfull_c^{(t)} \vert \boldsymbol{\mu}^{(t)}, \boldsymbol{\Lambda}^{(t)} \right)
\end{align}
with
\begin{align}\label{eq:proposal_params_direct}
    \begin{cases}
        \boldsymbol{\mu}^{(t)} =
        \paramfull^{(t-1)}
        - \frac{\epsilon}{2} \preconditioner^{(t-1)} \nabla g ( \paramfull^{(t-1)} )
        + \epsilon \correctiondrift^{(t-1)}
        , \\
        \boldsymbol{\Lambda}^{(t)} = \epsilon \preconditioner^{(t-1)}
        ,
    \end{cases}
\end{align}
where $\epsilon$ is a step size and $\correctiondrift^{(t-1)}$ is the additional drift term due to the position-dependent preconditioner~\cite{xifara_langevin_2014}.
In full generality, for all $i \in [\![1, ND ]\!]$,
%
\begin{align}
    \label{eq:def_correction_pmala}
    \gamma_i^{(t-1)}
    =
    \frac{1}{2}
    \sum_{j=1}^{N D} \frac{\partial G_{ij}^{(t-1)}}{\partial \theta_j^{(t-1)}}
    .
\end{align}
However, the RMSProp preconditioner is diagonal so that the sum in~\eqref{eq:def_correction_pmala} reduces to the $j=i$ term only.
Note that $\correctiondrift^{(t-1)}$ is defined from a differentiation with respect to iterate $\paramfull^{(t-1)}$ while the variance vector $\velocity$ in \eqref{eq:used_v_pmala} is defined from candidates.
Since all iterates start as candidates, let $j^{(t)}$ be the number of iterations since last accept: $j^{(t)} = \min \big\{ j \geq 0 \vert \paramfull^{(t)} = \paramfull^{(t-j)}_c \big\}$.
The correction terms $\gamma_i^{(t-1)}$ are then given by
\begin{align}
    \label{eq:used_correction_pmala}
    \gamma_i^{(t-1)}
    = -
    \frac{
        (1-\alpha)
        \alpha^{j^{(t-1)}}
        \left(
            \frac{\partial g}{\partial \paramelt{i}}
            \cdot
            \frac{\partial^2 g}{\partial \paramelt{i}^2}
        \right) \left(\paramfull^{(t-1)} \right)
    }{
        2 \sqrt{v_i^{(t-1)}} \left(\eta + \sqrt{v_i^{(t-1)}}\right)^2
    }
    ,
\end{align}
To compute $q(\paramfull^{(t-1)} \vert \paramfull^{(t)}_c) = \mathcal{N}\big( \paramfull^{(t-1)} \vert \boldsymbol{\mu}_c^{(t)}, \boldsymbol{\Lambda}_c^{(t)} \big)$,
one needs to update the variance $\velocity^{(t)}$ and preconditioner $\preconditioner^{(t)}$ and evaluate the candidate additional drift term $\correctiondrift_c^{(t)}$.
By definition, $j^{(t)}=0$ for candidates, so for all $i \in [\![1, ND ]\!]$,
\begin{align}\label{eq:used_correction_pmala_candidate}
    \gamma_{c,i}^{(t)} = - \frac{
        (1-\alpha)
        \left(
            \frac{\partial g}{\partial \paramelt{i}}
            \cdot
            \frac{\partial^2 g}{\partial \paramelt{i}^2}
        \right) \left(\paramfull^{(t-1)} \right)
    }{
        2 \sqrt{v_i^{(t)}} \left(\eta + \sqrt{v_i^{(t)}}\right)^2
    }
    .
\end{align}
%
The parameters $\boldsymbol{\mu}_c^{(t)}, \boldsymbol{\Lambda}_c^{(t)}$ are thus given by
\begin{align}\label{eq:proposal_params_candidate}
    \begin{cases}
        \boldsymbol{\mu}_c^{(t)} = \paramfull_c^{(t)}
        - \frac{\epsilon}{2} \preconditioner^{(t)} \nabla g \left( \paramfull_c^{(t)} \right)
        + \epsilon \correctiondrift_c^{(t)}
        , \\
        \boldsymbol{\Lambda}_c^{(t)} = \epsilon \preconditioner^{(t)}
        .
    \end{cases}
\end{align}
%
Algorithm~\ref{alg:pmala} describes the proposed PMALA kernel with RMSProp preconditioner.
It relies on three scalar parameters: a damping parameter $\eta$, an exponential decay rate $\alpha$ and a step size $\epsilon$.
The first two are generally set to $\eta = 10^{-5}$ and $\alpha = 0.99$~\cite{li_preconditioned_2016}.
The step size is chosen empirically.
MALA achieves optimal convergence rates with an acceptance rate equal to $0.574$ when the components of $\paramfull$ are independent~\cite{robert_monte_2004}.
Despite the interdependencies in the posterior distribution, we also set $\epsilon$ to obtain an average acceptance rate close to $0.574$, which yields good results in practice.



\begin{algorithm}[!t]
    \caption{PMALA kernel $\mathcal{K}_1$ at step $t$}
    \label{alg:pmala}
    \DontPrintSemicolon
    \KwIn{$\paramfull^{(t-1)}$,
    $\velocity^{(t-1)}$,
    $j^{(t-1)}$}
    \KwOut{$\paramfull^{(t)}$,
    $\velocity^{(t)}$,
    $j^{(t)}$}
    \vspace{2mm}
    \tcp{Propose candidate} 
    $\preconditioner^{(t-1)}$ and $\correctiondrift^{(t-1)}$
    \tcp*{using \eqref{eq:preconditioner}, \eqref{eq:used_correction_pmala}}
    $\boldsymbol{\mu}^{(t)}$ and $\boldsymbol{\Lambda}^{(t)}$
    \tcp*{using \eqref{eq:proposal_params_direct}}
    $\paramfull_c^{(t)} \sim \mathcal{N}(\boldsymbol{\mu}^{(t)}, \boldsymbol{\Lambda}^{(t)})$ \;
    \tcp{Accept or reject}
    $\velocity^{(t)}$, $\preconditioner^{(t)}$ and  $\correctiondrift_c^{(t)}$
    \tcp*{using \eqref{eq:used_v_pmala}, \eqref{eq:preconditioner}, \eqref{eq:used_correction_pmala_candidate}}
    $\boldsymbol{\mu}_c^{(t)}$, $\boldsymbol{\Lambda}_c^{(t)}$ and $\rho^{(t)}$
    \tcp*{using \eqref{eq:proposal_params_candidate}, \eqref{eq:accept_proba}}
    Draw $\zeta \sim \text{Unif}(0,1)$ \;
    \textbf{if} $\zeta \leq \rho^{(t)}$ \textbf{then} $\paramfull^{(t)} = \paramfull_c^{(t)}$, $j^{(t)} = 0$ \;
    \textbf{else} $\paramfull^{(t)} = \paramfull^{(t-1)}$, $j^{(t)} = j^{(t-1)} + 1$ \;
\end{algorithm}

\subsection{MTM transition kernel}

The non-log-concavity and potential multimodality of the posterior~\eqref{eq:posterior} is the second major difficulty to be addressed.
After a discussion on state-of-the-art methods, we will propose a MTM kernel.
In practice, samplers such as MH, MALA, HMC or even PMALA fail to explore the full distribution when modes are far away: they get stuck in one.
Alternative MCMC algorithms dedicated to multimodal distributions have been proposed in the literature.
Tempering-based samplers, e.g., the Equi-Energy Sampler~\cite{kou_equi-energy_2006} and the Adaptive Parallel Tempering Algorithm~\cite{miasojedow_adaptive_2013}, run parallel interacting Markov chains at different temperatures.
High temperature chains can navigate between modes and only one chain at low temperature is actually used for estimations.
%
%
Other methods consider an augmented distribution with a latent mode index and sample it with two kernels: a local kernel explores around a mode and a jump kernel permits jumps between modes.
Such methods include Darting MC (DMC)~\cite{andricioaei_smart_2001},
Jumping Adaptative Multimodal Sampler (JAMS)~\cite{pompe_framework_2020},
Regeneration Darting MC (RDMC)~\cite{ahn_distributed_2013}
and Wormhole HMC (WHMC)~\cite{lan_wormhole_2014}.
WHMC is a particular case of the Riemannian Manifold Hamiltonian Monte Carlo algorithm~\cite{girolami_riemann_2011}.
The metric of the corresponding manifold combines the standard Euclidean distance and a wormhole metric that shortens the distances between already identified modes, which simplifies transitions from one to another.
DMC and JAMS require a prior identification of the distribution modes by some optimization methods.
RDMC and WHMC allow running optimization methods in parallel to the sampler and update the distribution and the sampler parameters at random regeneration times~\cite{gilks_adaptive_1998}.
A more complete review of samplers dedicated to multimodal distributions is available in~\cite{pompe_framework_2020}.
Most of these methods are computationally very expensive or rely on optimization methods to identify the modes.
When $\approxf$ covers multiple decades and is non-linear, the posterior~\eqref{eq:posterior} has potentially many modes with only a few of significant weight in the distribution.
The identification of relevant modes with standard optimization methods is difficult.

We propose to use a Multiple-Try Metropolis (MTM)~\cite{martino_review_2018} kernel that can escape a local mode and explore other ones without any knowledge about the number, positions or variances of the modes.
%
%
Instead of sampling the whole vector $\paramfull \in \R^{ND}$ at once, it uses a Gibbs sampler to decompose it into $N$ individual $\paramvect{n}$.
For each conditional distribution, it harnesses an Independent Multiple-Try Metropolis (I-MTM) approach~\cite{liu_multiple-try_2021,martino_flexibility_2013,martino_review_2018}.
This method generates $K \geq 1$ candidates $(\paramvect{n}^{(k)})_{k=1}^K$ independently of $\paramvect{n}^{(t-1)}$.
This divide-to-conquer approach permits considering $N$ conditional distributions $\pi \big(\paramvect{n} \vert \obsvect{n}, \paramfull^{(t-1)}_{\setminus n} \big)$ of small dimension, where $\paramfull^{(t-1)}_{\setminus n} = \big( \paramvect{1}^{(t-1)}, \ldots,  \paramvect{n-1}^{(t-1)}, \paramvect{n+1}^{(t-1)}, \ldots, \paramvect{N}^{(t-1)} \big)$. 
%
%
%
%
%
%
Candidates are sampled from a proposal distribution $q \big(\paramvect{n} \vert \paramfull^{(t-1)}_{\setminus n} \big)$ that should be permissive enough to generate candidates in all modes of $\pi \big(\paramvect{n} \vert \obsvect{n}, \paramfull^{(t-1)}_{\setminus n} \big)$.
%
Then, using an importance weight function $w$
%
\begin{align}\label{eq:importance_weight_MTM}
    w \big(\paramvect{n}^{(k)} \big) & = \frac{
        \pi \big(\paramvect{n}^{(k)} \vert \obsvect{n}, \paramfull^{(t-1)}_{\setminus n}\big)
    }{
        q \big(\paramvect{n}^{(k)} \vert \paramfull^{(t-1)}_{\setminus n}\big)
    },
\end{align}
one candidate is selected using a categorical distribution with selection probability $w_k$ for candidate $k$
\begin{align}\label{eq:selection_proba_MTM}
    w_k
    =
    \frac{w \big(\paramvect{n}^{(k)} \big)}{\sum_{j=1}^K w \big(\paramvect{n}^{(j)} \big)}
    .
\end{align}
The MH step is then performed with the selected candidate $i$ and the generalized acceptance probability~\cite{liu_multiple-try_2021,martino_review_2018}
\begin{equation}
    \label{eq:generalized_accept_proba}
    r^{(t)} =
        1
        \wedge
        \frac{
            w \big(\paramvect{n}^{(i)} \big) + \sum_{j=1,j \neq i}^K w \big(\paramvect{n}^{(j)} \big)
        }{
            w \big(\paramvect{n}^{(t-1)} \big) + \sum_{j=1,j \neq i}^K w \big(\paramvect{n}^{(j)} \big)
        }
        .
\end{equation}
%
Algorithm~\ref{alg:mtm} summarizes the MTM sampler.
Note that due to the Gibbs approach, it updates one component at a time and returns the result of all updates.
A succession of intermediate updates $\big(\paramfull^{(t - 1 + n/N)} \big)_{n=1}^N$ is therefore introduced.
This transition kernel relies on the choice of the proposal distribution $q$ and on the number of candidates $K$ generated at each step.
This parameter is chosen as a trade-off between computational intensity and average acceptance probability: the higher $K$, the higher the acceptance probability, the mixing capability but also the computational cost.

In image inverse problems, many common spatial priors are based on a local operator such as the image gradient or Laplacian.
In such cases, many components $\paramvect{n}$ are conditionally independent.
They can be sampled in parallel using a Chromatic Gibbs sampler~\cite{gonzalez_parallel_2011}, which can significantly speed up computations. 

\begin{algorithm}[!t]
    \caption{MTM kernel $\mathcal{K}_2$ at step $t$}
    \label{alg:mtm}
    %
    \DontPrintSemicolon
    \KwIn{$\paramfull^{(t-1)}$}
    \KwOut{$\paramfull^{(t)}$}
    \vspace{1mm}
    \For{$n = 1$ \KwTo $N$}{
        \tcp{Propose candidates, select one}
        $\paramvect{n}^{(k)} \sim q \Big(\paramvect{n} \Big\vert \paramfull^{\left(t-1 + \frac{n-1}{N}\right)}_{\setminus n} \Big)$
        for $k = 1$ to $K$ \;
        %
        %
        $w \big(\paramvect{n}^{(k)} \big)$
        for $k = 1$ to $K$
        \tcp*{using \eqref{eq:importance_weight_MTM}}
        $w_k$
        for $k = 1$ to $K$
        \tcp*{using \eqref{eq:selection_proba_MTM}}
        $i \sim \text{Cat}(w_1, \cdots, w_K)$ \;
        \tcp{Accept or reject}
        $r_n^{(t)}$
        \tcp*{using \eqref{eq:generalized_accept_proba}}
        Draw $\zeta \sim \text{Unif}(0,1)$ \;
        \uIf{$\zeta \leq r_n^{(t)}$}{
            $\paramvect{n}^{\left(t - 1 + \frac{n}{N}\right)} = \paramvect{n}^{(i)}$,
            $\paramfull^{\left(t - 1 + \frac{n}{N}\right)}_{\setminus n} = \paramfull^{\left(t - 1 + \frac{n-1}{N}\right)}_{\setminus n}$
        }
        \textbf{else} $\paramfull^{\left(t - 1 + \frac{n}{N}\right)} = \paramfull^{\left(t - 1 + \frac{n-1}{N}\right)}$ \;
    }
\end{algorithm}

\subsection{Proposed sampler and implementation details}

To combine a good local exploration of modes as well as jumps between modes, the proposed kernel mixes the PMALA and MTM transition kernels above.
At every step $t$, the MTM kernel is selected with probability $p$, and the PMALA kernel with probability $1-p$.
Since the MTM kernel divides the parameter space in $N$ $D$-dimensional subspaces, the PMALA global integer $j^{(t)} \geq 0$ is replaced by a vector $\boldsymbol{j}^{(t)} \in \mathbb{N}^N$, where $j_n^{(t)}$ counts the number of steps since last acceptance for component $\paramvect{n}$.
When a component $\paramvect{n}$ is accepted by the MTM kernel, the counter $j_n$ is reset to $0$ and the variance component $\boldsymbol{v}_n \in \R^D$ is updated as in~\eqref{eq:used_v_pmala} with $\frac{\partial g}{\partial \paramvect{n}} \big( \paramfull^{(t)} \big)$.

Algorithm~\ref{alg:final} reports the complete proposed sampler.
Similarly to RDMC and WHMC, the proposed sampler mixes a kernel dedicated to local exploration -- PMALA -- and another to jump between modes -- MTM.
The decomposition of the parameter space into $N$ $D$-dimensional subspaces makes the sampler much simpler than previous approaches.
%
%
It will perform well in structured problems that allow such decomposition, e.g., images and graphs, and poorly in high-dimensional problems that do not, e.g., Gaussian Mixtures over the full space.

Regarding theoretical properties, the PMALA kernel satisfies the detailed balance property -- from~\cite[theorem 7.2]{robert_monte_2004} -- and produces ergodic Markov chains -- from~\cite[corollary 7.5]{robert_monte_2004}).
The proposed MTM kernel is a Metropolis-within-Gibbs algorithm with propositions independent to the current location and with multiple candidates $K$.
In the particular case where $K=1$, it satisfies the detailed balance property and produces uniformly ergodic Markov Chains -- from~\cite[theorem 7]{jones_convergence_2014}.
Using $K > 1$ candidates in a Multiple-Try Metropolis framework maintains detailed balance and ergodicity~\cite{martino_review_2018}.
As a mixture of kernels having the same stationary distribution, the proposed kernel also admits the posterior as a stationary distribution -- from~\cite[chapter 10]{robert_monte_2004}.
As the MTM kernel produces uniformly ergodic Markov chains, so does the proposed mixture kernel -- from~\cite[proposition 10.20]{robert_monte_2004}).
These results of convergence towards the posterior are mostly asymptotic and also hold for simpler algorithms such as Random Walk Metropolis-Hastings~\cite{roberts_geometric_1996}.
A comparative theoretical study of non-asymptotic properties that could demonstrate a faster convergence of the proposed sampler is beyond the scope of this paper.
However, empirical results show that the proposed sampler yields state-of-the-art performance on multimodal distributions in low-dimensional settings, or higher dimensional applications with relevant low-dimensional conditional distributions.

\begin{algorithm}[!t]
    \caption{Proposed sampler: PMALA and MTM}
    \label{alg:final}
    \DontPrintSemicolon
    \KwIn{number of iterations $T$,
    starting point $\paramfull^{(0)}$}
    \KwOut{Markov chain $\{ \paramfull^{(t)} \}_{t=1}^T$}
    \vspace{2mm}
    Initialize $v_{nd}^{(0)} = \big[ \frac{\partial g}{\partial \paramelt{nd}}  \big(\paramfull^{(0)} \big) \big]^2$ for all $n$ and $d$ \;
    Initialize $\boldsymbol{j}^{(0)} = \boldsymbol{0}_{N}$ \;
    \For{$t = 1$ \KwTo $T$}{
        Draw $\zeta \sim \text{Unif}(0,1)$ \;
        \uIf(\tcp*[h]{PMALA kernel (Algo.~\ref{alg:pmala})}){$\zeta > p$}{
            $\paramfull^{(t)}, \velocity^{(t)}, \boldsymbol{j}^{(t)} = \mathcal{K}_1 \big( \paramfull^{(t-1)}, \velocity^{(t-1)}, \boldsymbol{j}^{(t-1)} \big)$ \;
        }
        \Else(\tcp*[h]{MTM kernel (Algo.~\ref{alg:mtm})}){
            $\paramfull^{(t)} = \mathcal{K}_2 \big( \paramfull^{(t-1)} \big)$ \;
            \tcp{Update PMALA parameters}
            \For{$n = 1$ \KwTo $N$}{
                \If($\forall d, $){candidate for $\paramvect{n}$ was accepted}{
                    $v_{nd}^{(t)} = \alpha v_{nd}^{(t-1)} + (1 - \alpha)  \big[ \frac{\partial \neglogpost }{\partial \paramelt{nd}} ( \paramfull^{(t)} ) \big]^2$, \;
                    \vspace{1mm}
                    $j_n^{(t)} = 0$,
                    $\velocity_{\setminus n}^{(t)} = \velocity_{\setminus n}^{(t-1)}$,
                    $\boldsymbol{j}_{\setminus n}^{(t)} = \boldsymbol{j}_{\setminus n}^{(t-1)}$ \;
                }
            }
        }
    }
\end{algorithm}

\section{Numerical experiments}\label{section:experiments}

The performance of the proposed method is evaluated on three examples of increasing complexity:
(i)~sampling a 2D-Gaussian mixture model with unknown modes,
(ii)~the sensor localization problem~\cite{ihler_nonparametric_2005},
and (iii)~a synthetic astrophysical inverse problem inspired from~\cite{wu_constraining_2018}.
The first two examples focus on the ability of the sampler to efficiently explore multimodal distributions.
The astrophysical inverse problem combines all the challenges addressed in Sections~\ref{section:model} and~\ref{section:mcmc}: censorship, mixture of noises, forward model spanning multiple decades and multiple local minima.

\subsection{Gaussian mixture model}

A two-dimensional Gaussian mixture model (GMM) restricted to the square $\fvalidity = [-15, 15]^2$ is considered.
This simple multimodal distribution, shown on Fig.~\ref{fig:GMM} (top left), is set to contain 15 modes $(\boldsymbol{\mu}_i, \boldsymbol{\Sigma}_i)$. 
It will demonstrate the ability of the proposed sampler to jump between modes.
For simplicity, all the modes have an equal weight in the mixture
%
\begin{align}
    \pi(\paramfull) \propto \left[
        \sum_{i=1}^{15}
        \mathcal{N} \left(
            \paramfull \vert \boldsymbol{\mu}_i, \boldsymbol{\Sigma}_i
        \right)
    \right]
    \exp \left( - \delta \, \tilde{\iota}_{\fvalidity}(\paramfull) \right),
\end{align}
with $\delta = 10^4$.
No natural structure decomposition exists for a GMM since each observation consists of $N = 1$ point only in dimension $D = 2$.
%
A Markov chain composed of $10\,000$ samples is considered, including $100$ burn-in samples.
To illustrate the role of each of the two kernels in the proposed sampler, two different values are considered for the probability of selecting the MTM kernel: $p = 0.1$ or $p = 0.9$.
The number of candidates of the MTM kernel is set to $K = 50$, and the proposal distribution $q$ is the smooth uniform prior on $\mathcal{C}$ (see appendix~\ref{section:sampling_smmoth_indicator}).
The MTM candidates weights $w \big( \paramvect{n}^{(k)}\big)$ in~\eqref{eq:importance_weight_MTM} are then equal to the likelihood term, i.e., the sum of Gaussian pdfs.
The default values $\alpha = 0.99$ and $\eta = 10^{-5}$ are considered for the exponential decay and damping factor of the PMALA kernel~\cite{li_preconditioned_2016}, and its step size is set to $\epsilon =  0.5$.
%
The proposed approach is compared to the state-of-the-art Wormhole Hamiltonian Monte Carlo sampler (WHMC)~\cite{lan_wormhole_2014}, using the same number of samples.
Note that WHMC needs the prior knowledge of mode positions $(\boldsymbol{\mu}_i)_{1 \leq i \leq 15}$, while the proposed kernel does not.

Fig.~\ref{fig:GMM} shows the 2D histograms obtained with the three samplers: WHMC and the proposed ones with either $p=0.1$ or $p=0.9$.
The three Markov chains efficiently explore all the modes, and their local dispersion obeys the covariance structures equally well. 
%
Table~\ref{tab:results_gmm} compares their effective sample sizes (ESS)~\cite{robert_monte_2004} and biases.
%
When $p = 0.9$,
the proposed sampler achieves better performances than WHMC, despite the absence of information about the position of the modes $\boldsymbol{\mu}_i$.
The high ESS values result from the 85\% acceptance rate of the MTM kernel for $K = 50$.
However, the MTM kernel with a fixed number of candidates $K$ would not scale up to much higher dimensions.
The probability to jump between modes is proportional to the volume of the high probability regions compared to the volume of $\fvalidity$, and thus decreases exponentially with the dimension of the problem.
The proposed sampler would therefore fail to reach isolated modes in a high-dimensional GMM, whereas WHMC would succeed to do so by exploiting its additional information about the modes.
However, the proposed approach focuses on scenarios where the parameter space can be partitioned into a collection of $N$ subspaces of limited dimension $D$, typically $D \lesssim 10$.
The MTM kernel thus remains out of reach from the curse of dimension thanks to the structure of the problem.
As in this simple GMM example, the proposed sampler can then outperform WHMC, even without any prior information on the modes of a multimodal distribution.

\begin{table}[!t]
    \begin{center}
    \caption{Samplers comparison on 2D-GMM.}
    \label{tab:results_gmm}
    \begin{tabular}{|c|c|c|c|}
    \hline
    \multirow{2}{*}{MCMC sampler}
    & Bias
    & \multicolumn{2}{c|}{ESS} \\
    \cline{3-4}
    & $\left\Vert \mathbb{E}[\paramfull] - \paramfull^* \right\Vert $ & $\paramelt{1,1}$ & $\paramelt{1,2}$ \\
    \hline
    WHMC~\cite{lan_wormhole_2014} &  $1.28 \cdot 10^{-1}$ & 2\,753 & 2\,993 \\
    \hline
    Proposed, $p=0.1$  & $7.02 \cdot 10^{-1}$ & 395 & 444 \\
    Proposed, $p=0.9$ & $\boldsymbol{4.61 \cdot 10^{-2}}$ & \textbf{6\,157} & \textbf{5\,780} \\
    \hline
    \end{tabular}
    \end{center}
\end{table}

\begin{figure}[!t]
    \centering
    \hspace{-2mm}
    \subfloat{
        \includegraphics[height=3.2cm, width=0.205\textwidth]{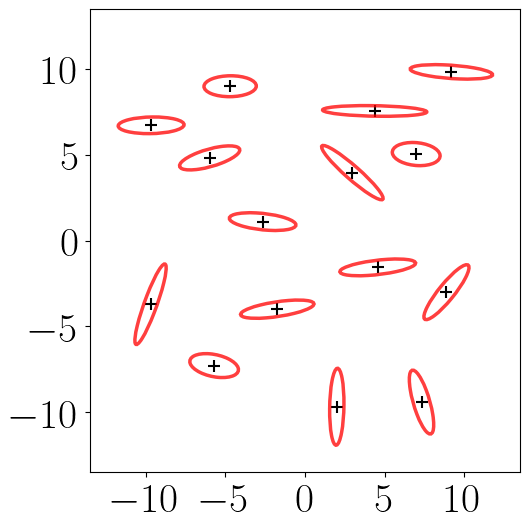}
    }
    \hspace{5mm}
    \subfloat{
        \includegraphics[height=3.2cm, width=0.24\textwidth]{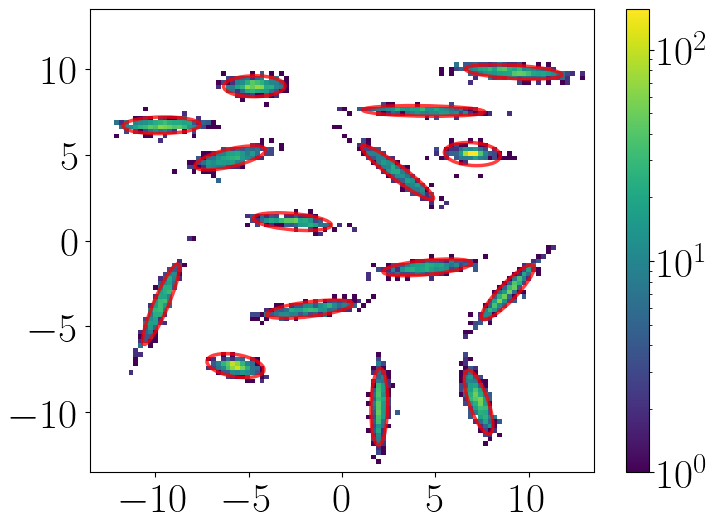}
    } \\
    \subfloat{
        \includegraphics[height=3.2cm, width=0.24\textwidth]{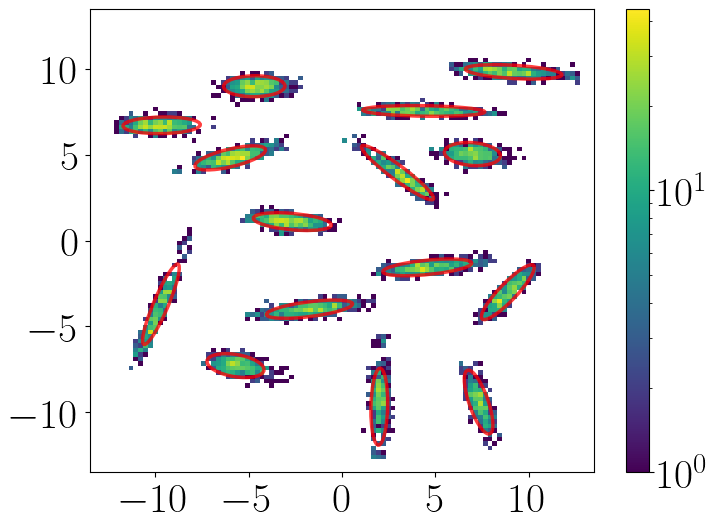}
    }
    \subfloat{
        \includegraphics[height=3.2cm, width=0.24\textwidth]{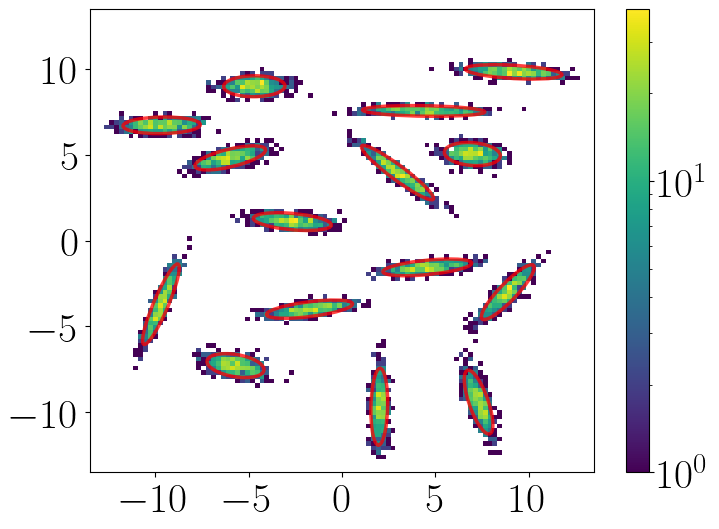}
    }
    \vspace{-2mm}
    \caption{
        Sampling results on GMM for
        WHMC (top right),
        the proposed kernel with $p = 0.1$ (bottom left)
        and $p = 0.9$ (bottom right).
        %
        The red ellipses show the probability level at $2\sigma$.
        All histograms are in logarithmic norm.
    }
    \label{fig:GMM}
\end{figure}

\subsection{Sensor localization}

The sensor localization problem introduced in~\cite{ihler_nonparametric_2005} is a common test case in multimodal sampling, e.g., in~\cite{ahn_distributed_2013, lan_wormhole_2014, pompe_framework_2020}.
Three sensors have known locations and will serve as a reference to avoid ambiguities with respect to translation, rotation and negation.
The goal is to estimate the unknown positions $\paramfull \in \R^{N D}$ of $N = 8$ sensors in dimension $D = 2$.
The observation matrix $\obsfull \in \R^{N L}$ collects noisy and partially censored pairwise distances between sensors located in a square $[0,1]^2$, where $L = N + 3$ is the total number of sensors.
%
The distance to sensor $\ell$ feeds channel $\ell$, so that the forward model is $\truefell(\paramvect{n})= \Vert \paramvect{n} - \paramvect{\ell} \Vert$.
Note that only $N+2$ distances will really be used since $\truefell(\paramvect{\ell})=0$, and that we set $y_{n,\ell} = 0$ by convention.
The probability of communication from sensor $\ell  \in [\![1, L]\!]$ to sensor $n \in [\![1, N]\!]$ is set to $\exp \left\{-\frac{\truefell(\paramvect{n})^2}{2 R^2} \right\}$ with $R = 0.3$.
%
In absence of communication, the observation is censored, which is encoded by the binary latent variable $c_{n,\ell} = 1$.
Otherwise, $c_{n,\ell}=0$ when the observation occurs and is corrupted by a white Gaussian noise
\begin{equation}
    y_{n,\ell} =
    \truefell(\paramvect{n})
    + \epsilon_{n,\ell},
    \quad \text{with }
    \epsilon_{n,\ell} \sim \mathcal{N}(0, \sigma_{\epsilon}^2)
    ,
\end{equation}
with $\sigma_\epsilon = 0.02$, leading to
%
%
%
\begin{align}
    -  \log \pi & \left( \obsfull \vert \paramfull \right)
    =
    \sum_{n=1}^N
    \sum_{\ell=1}^{L}
    (1 - c_{n,\ell}) \left[
        \frac{(\truefell(\paramvect{n}) - y_{n,\ell})^2}{2 \sigma_\epsilon^2}
    \right. \\
    & \left.
        + \frac{\truefell(\paramvect{n})^2}{2 R^2}
    \right]
    + c_{n,\ell} \log \left[
        1 - \exp \left( - \frac{\truefell(\paramvect{n})^2}{2 R^2} \right)
    \right] \nonumber
    ,
\end{align}
%
%
%
%
The smoothed uniform prior on the square $\fvalidity = [-0.35, 1.2]^2$ is used as a prior on the location of each sensor.
The corresponding penalty parameter $\delta$ introduced in~\eqref{eq:overall_prior} is set to $10^4$.
This prior is non-informative enough to match the results shown in~\cite{ahn_distributed_2013,lan_wormhole_2014,pompe_framework_2020}.
The proposed sampler is compared to both Regeneration Darting Monte Carlo (RDMC)~\cite{ahn_distributed_2013} and WHMC.
A Markov chain of size $30\,000$ is generated by each algorithm, including $5\,000$ burn-in samples.
The parameters of the PMALA kernel are set to $\alpha = 0.99$, $\eta = 10^{-5}$ and $\epsilon = 3 \times 10^{-3}$.
The MTM kernel is selected with $p=0.1$ or $p=0.9$.
Its proposal distribution $q$ is set to the smooth uniform prior on $\fvalidity$.
%
For each sensor, the high probability regions are small compared to $\fvalidity$.
To obtain high acceptance rates for the MTM kernel, the number of candidates is set to $K = 1\,000$.
Better proposal distributions can be obtained for this specific problem, which is beyond the scope of this experiment.

Fig.~\ref{fig:sensor_loc} shows the marginal distributions of each sensor position.
The four samplers identified the same modes.
Table~\ref{tab:ess_sensor_loc} compares the samplers in terms of ESS.
With $p = 0.9$, the proposed sampler yields better mixing capability than WHMC and RDMC.
%
This is due to the partition of the $ND = 16$-dimensional problem into $N=8$ simpler $D=2$-dimensional problems.
This divide-to-conquer strategy exploits the problem structure to fight the curse of dimension.

\begin{table}[!t]
    \caption{Effective Sample Size (ESS) on the sensor localization problem.}
    \label{tab:ess_sensor_loc}
    \begin{center}
    \begin{tabular}{|c|c|c|c|}
    \hline
    \multirow{2}{*}{MCMC sampler} & \multicolumn{3}{c|}{ESS} \\
    \cline{2-4}
     & min & mean & max\\
    \hline
    WHMC~\cite{lan_wormhole_2014} & 29 & 1\,026 & 5\,753\\
    RDMC~\cite{ahn_distributed_2013} & 168 & 3\,354 & 11\,192\\
    \hline
    Proposed, $p=0.1$ & 29 & 329 & 1\,235\\
    Proposed, $p=0.9$ & \textbf{299} & \textbf{3\,561} & \textbf{16\,789}\\
    \hline
    \end{tabular}
    \end{center}
\end{table}

\begin{figure}[!t]
    \centering
    \subfloat{
        \includegraphics[height=3.5cm, width=0.23\textwidth]{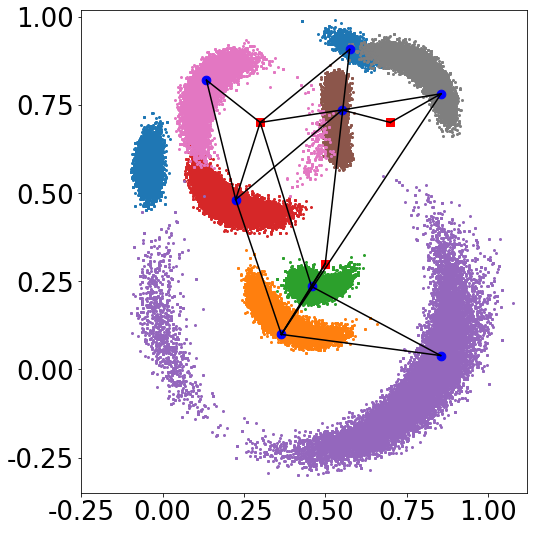}
    }
    \,
    \subfloat{
        \includegraphics[height=3.5cm, width=0.23\textwidth]{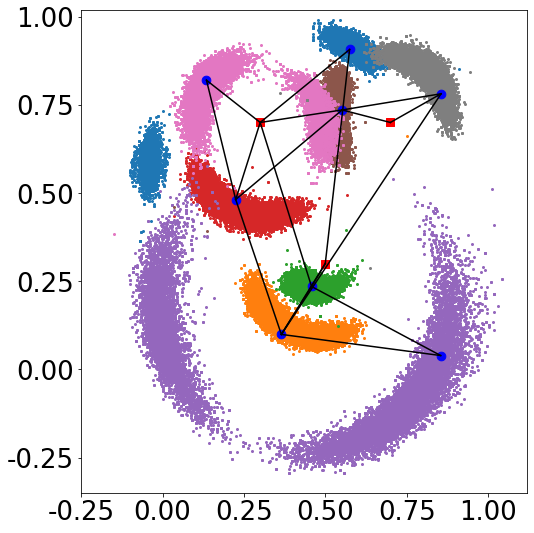}
    } \\
    \subfloat{
        \includegraphics[height=3.5cm, width=0.23\textwidth]{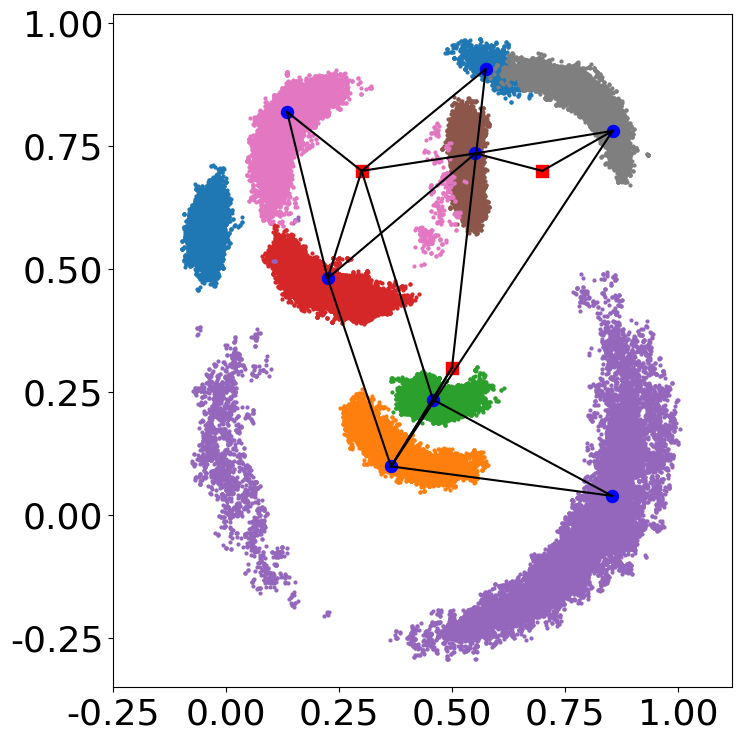}
    }
    \,
    \subfloat{
        \includegraphics[height=3.5cm, width=0.23\textwidth]{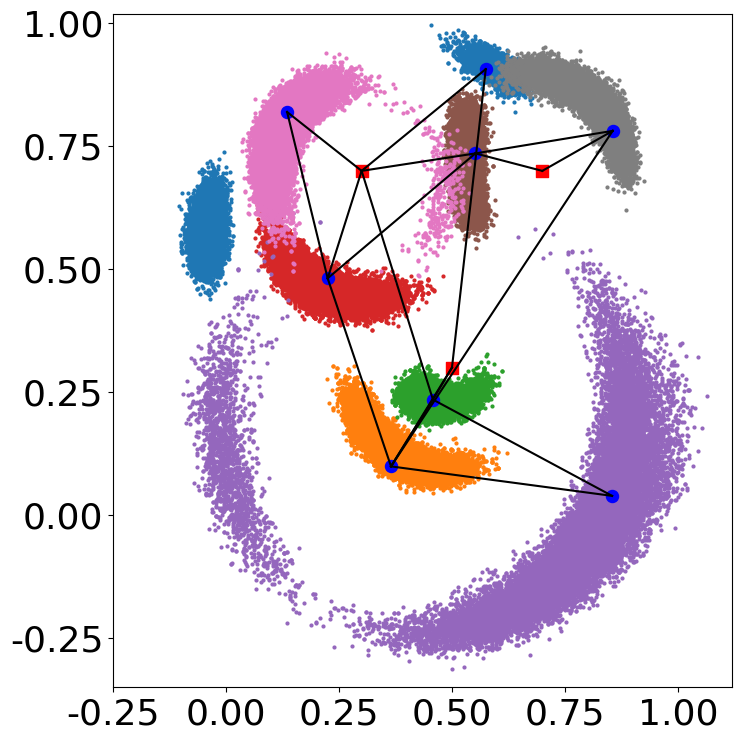}
    }
    \caption{
        Marginal distributions of the sensors positions for
        RDMC (top left),
        WHMC (top right),
        proposed with $p = 0.1$ (bottom left)
        and with $p = 0.9$ (bottom right).
        The graph shows the true position of all sensors.
        The sensors with a known position are in red and those whose position is inferred are in blue.
        The edges of the graph indicate which pairs of sensors are observed.
    }
    \label{fig:sensor_loc}
\end{figure}

\subsection{Realistic astrophysical synthetic data}\label{subsec:astro_ex}

The overall approach is now applied to a synthetic yet realistic inverse problem from astrophysics~\cite{wu_constraining_2018,joblin_structure_2018}.
The goal is to reconstruct maps of physical parameters of a molecular cloud from radio wave multispectral intensity maps.
Each observation map contains $N = 4\,096$ pixels.
Each pixel is associated to $D = 4$ physical parameters $\varphibf = (\kappa, \, P_{th}, \, G_0, \, A_V)$, so that the aim is to infer a set of parameters $\Phibf = (\varphibf_n)_{n=1}^N$ in dimension $N\times D = 16\,536$.
The parameter $\kappa$ is a nuisance parameter related to the conditions of observations. 
It's ground truth value is set to 1 over the whole map.
The main parameters of interest are the thermal pressure $P_{th}$, the intensity of a UV radiative field $G_0$ and the visual extinction $A_V$, related to the cloud depth along the line of sight.
%
The ground truth parameters $\Phibf^*$ are chosen according to a plausible astrophysical scenario~\cite{pety_anatomy_2016}.
%
The physics of the system is encoded within the Meudon PDR code~\cite{le_petit_model_2006}, a large numerical simulator.
This forward model features many properties that make inference difficult: it is a non-linear model that yields a multimodal posterior distribution, and the amplitude of observations as well as parameters $\varphibf$ span several decades.
%
%
%
A discrete grid of values $ \{ (\varphibf[g], \boldsymbol{f}[g]), g \in \mathcal{G} \}$ is used to define a normalization process as well as the reduced model.
To work with similar scales, the set of estimated parameters $\paramfull$ will correspond to normalized values $\paramvect{}$ of $\log\varphibf$ with respect to empirical averages and variances of $\{ \log \varphibf[g], g \in \mathcal{G} \}$.
To avoid repeated expensive evaluations, the forward model $\truef$ is reduced to an approximate model $\approxf{}$, as in~\eqref{eq:approxfell}.
%
For each line $\ell$, a polynomial approximation $\approxPell$ of degree 6 is trained on collection $\{ (\paramvect{}[g], \log \truefell[g]), g \in \mathcal{G} \}$. 
The approximation quality of the resulting $\approxf{}$ will be considered of sufficient quality to replace exact simulations everywhere. 
It is used to generate observation maps of $L = 10$ emission lines.
For each line $\ell$, $\approxfell$ ranges from $10^{-18}$ to $10^{-2}$.
These maps are deteriorated by additive noise, multiplicative noise and censorship following the observation model~\eqref{eq:model}.
The standard deviation of the multiplicative noise is set to $\sigma_m = \log(1.1)$, which roughly represents a $10$\% alteration in average.
For the additive noise, $\sigma_a = 1.38715 \cdot 10^{-10}$ so that the Signal-to-Noise Ratio~(SNR) varies between $-81$ and $79$ dB.
Observations $y_{n,\ell}$ range from about $10^{-10}$ to $10^{-2}$.
The censorship level is set to $\omega = 3 \sigma_a$.
Fig.~\ref{fig:astro_obs} shows the observation maps of two lines and the spatial distribution of censorship importance.

\begin{figure}[!t]
    \centering
    \subfloat{
        \includegraphics[height=2.4cm, width=0.328\linewidth]{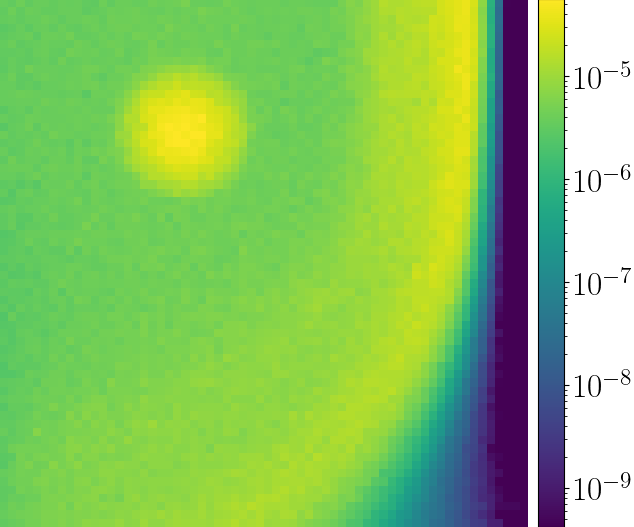}}
    \hfill%
    \subfloat{
        \includegraphics[height=2.4cm, width=0.328\linewidth]{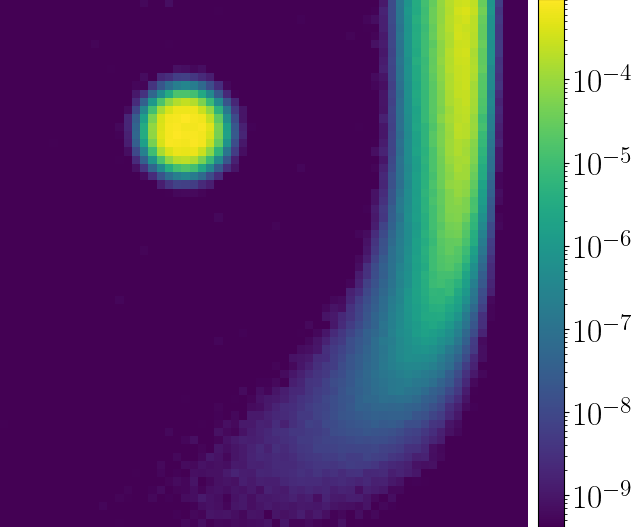}}
    \hfill%
    \subfloat{
        \includegraphics[height=2.4cm, width=0.328\linewidth]{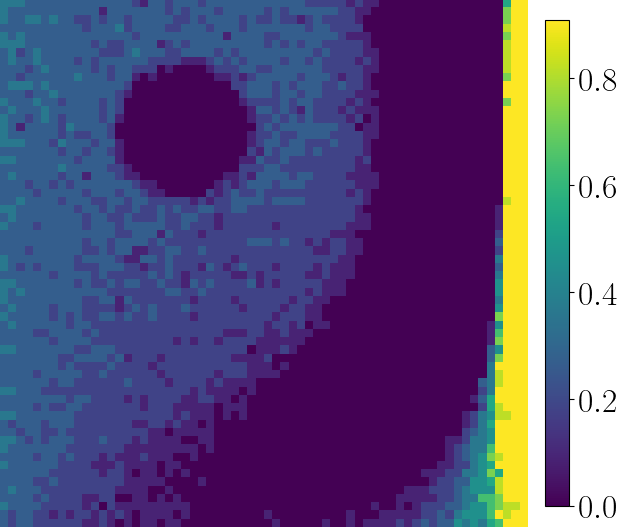}}
    \caption{
        Some observation maps of the astrophysical experiment.
        From left to right:
        line $\ell = 1$,
        line $\ell = 10$,
        proportion of censored lines per pixel.
    }
    \label{fig:astro_obs}
\end{figure}

\begin{figure}
    \centering
    \subfloat{
    \includegraphics[height=2.5cm,width=0.30\linewidth]{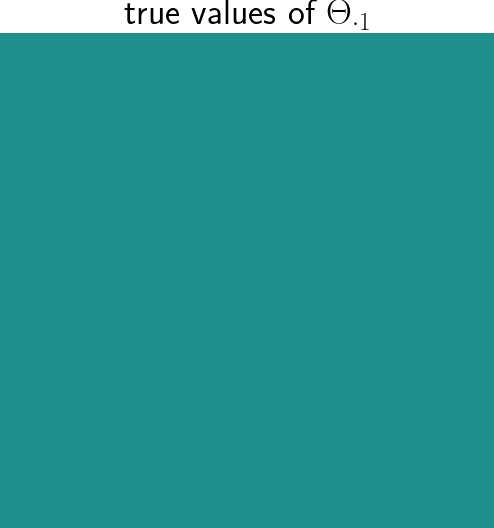}}
    \,
\subfloat{
     \includegraphics[height=2.5cm,width=0.33\linewidth]{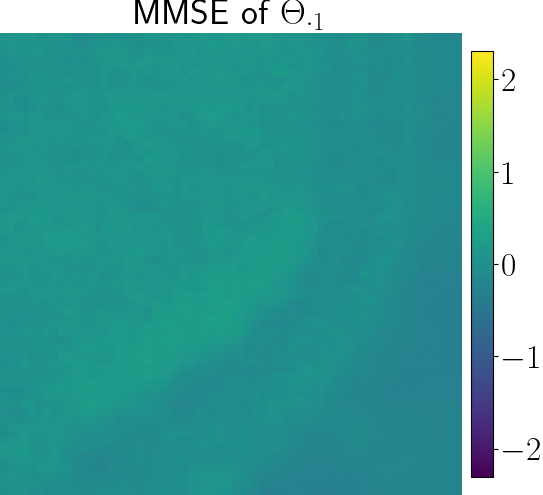}}
 \,
 \subfloat{
     \includegraphics[height=2.5cm,width=0.33\linewidth]{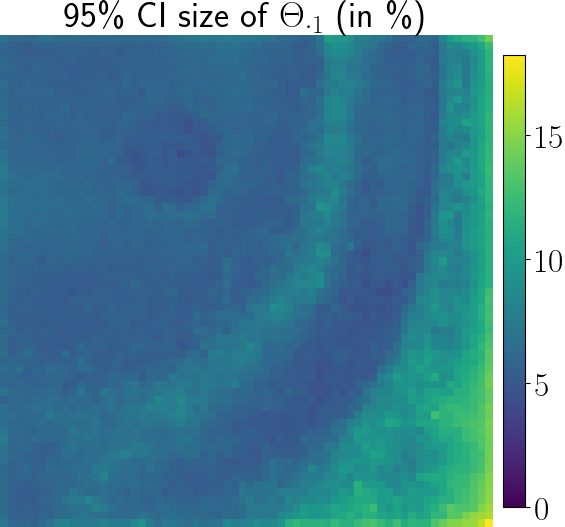}} \\
 \vspace{-2mm}
 \subfloat{
       \includegraphics[height=2.5cm,width=0.30\linewidth]{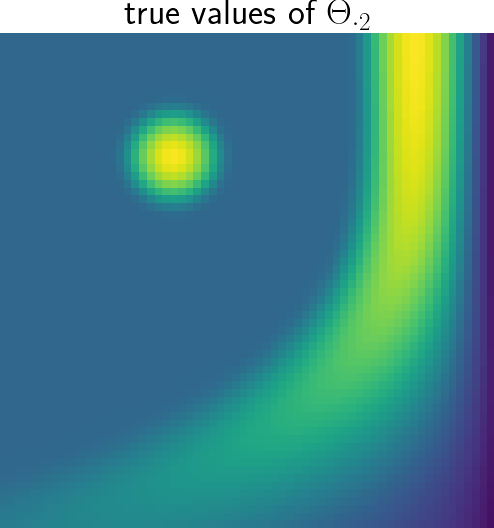}}
    \,
  \subfloat{
        \includegraphics[height=2.5cm,width=0.325\linewidth]{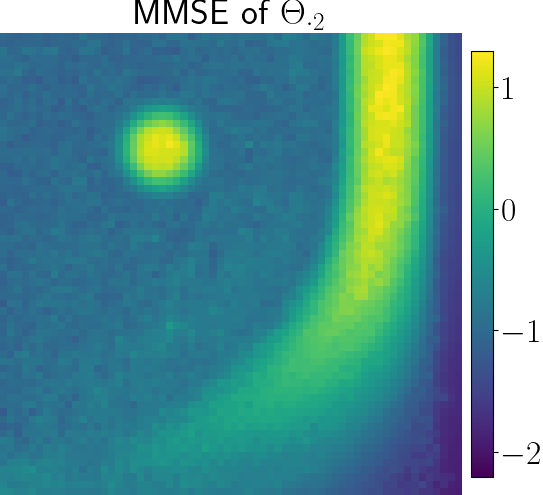}}
    \,
    \subfloat{
        \includegraphics[height=2.5cm,width=0.325\linewidth]{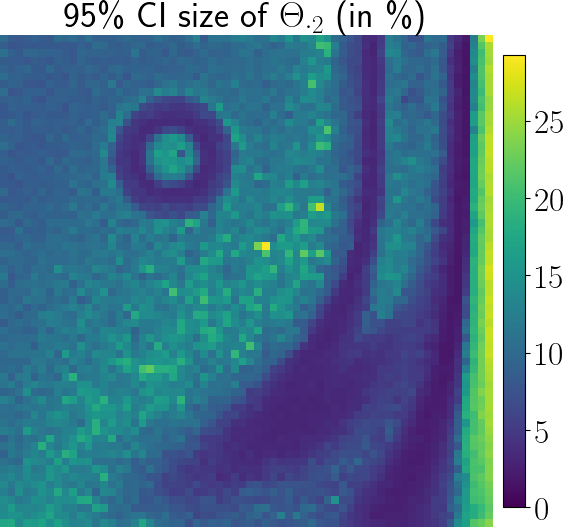}}\\
    \vspace{-2mm}
    \subfloat{
       \includegraphics[height=2.5cm,width=0.30\linewidth]{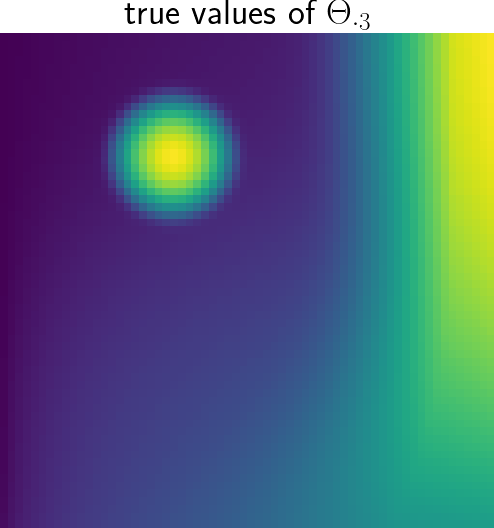}}
    \,
  \subfloat{
        \includegraphics[height=2.5cm,width=0.325\linewidth]{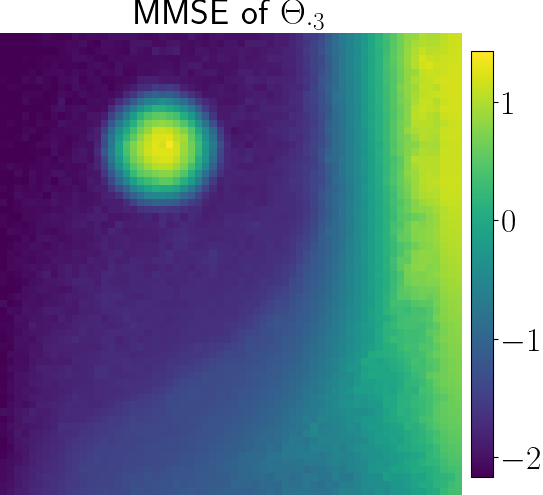}}
    \,
    \subfloat{
        \includegraphics[height=2.5cm,width=0.325\linewidth]{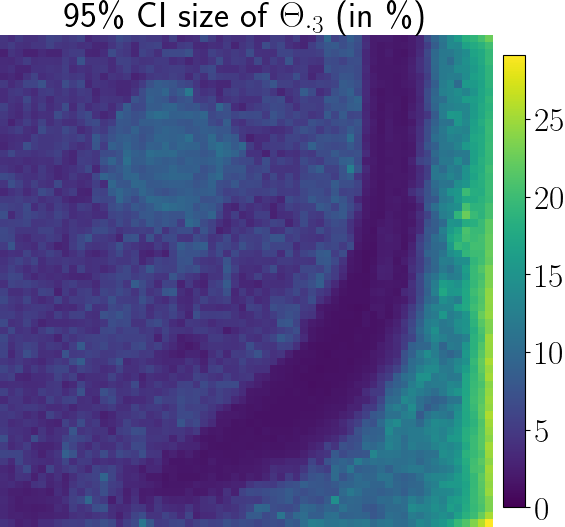}} \\
    \vspace{-2mm}
    \subfloat{
        \includegraphics[height=2.5cm,width=0.30\linewidth]{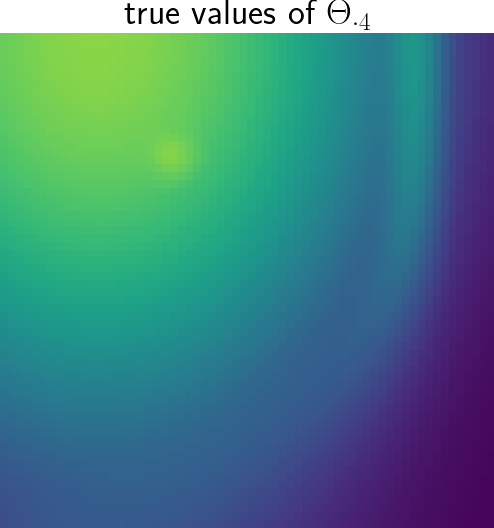}}
    \,
    \subfloat{
        \includegraphics[height=2.5cm,width=0.325\linewidth]{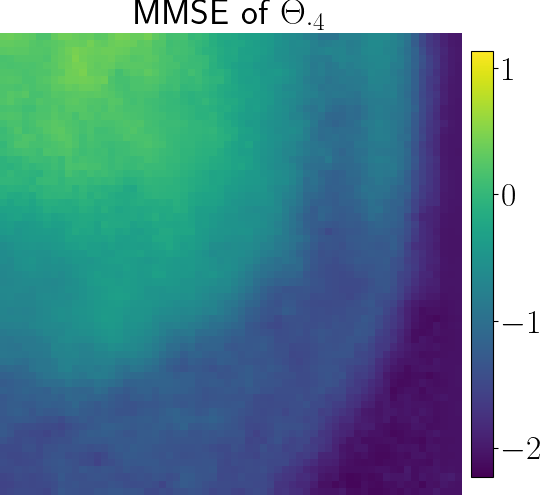}}
    \,
    \subfloat{
        \includegraphics[height=2.5cm,width=0.325\linewidth]{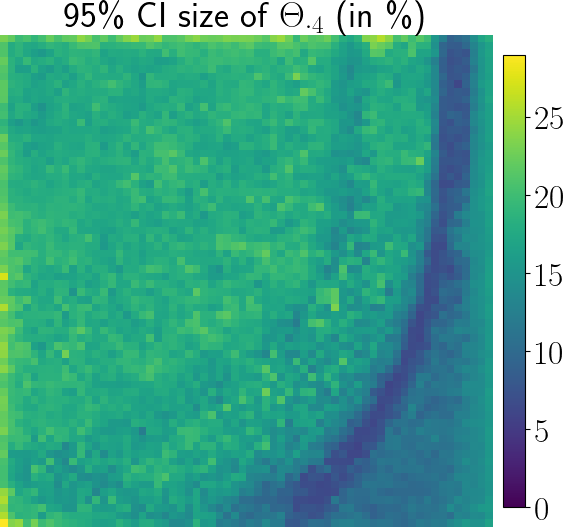}}
    \caption{
        Inference results:
        (left) ground truth $\paramfull^*$;
        (middle) MMSE estimate from the proposed transition kernel;
        (right) size of the $95$\% credibility intervals (CI) in \% of the size of the validity intervals.
    }
    \label{fig:astro_results}
\end{figure}

The likelihood approximation is obtained as indicated in Section~\ref{ssub:noise_approx}, and its parameters $\aell$ are adjusted as described in Appendix~\ref{section:llh_param_optim}.
The validity set $\fvalidity$ of physical parameters is set as in~\cite{wu_constraining_2018}, and the penalty parameter $\delta$ of the smooth uniform prior is set to $10^4$.
%
Given the smoothness of the true maps, for each $d$ the chosen spatial regularizer $h$ is taken as
\begin{align}\label{eq:spatial_prior}
    h(\paramfull_{\cdot d})
    =
    \Vert \Delta \paramfull_{\cdot d} \Vert^2_2
    =
    \sum_{n = 1}^N \sum_{i \in V_n} (\paramelt{n,d} - \paramelt{i,d})^2
    ,
\end{align}
where $\Delta$ is the discrete (5-point based) 2D Laplacian operator,
and $V_n$ is the set of neighbors of pixel $n$ induced by $\Delta$.
The hyperparameter $\reguweight$ from~\eqref{eq:overall_prior} is fixed to $\reguweight = (10, 2, 3, 4)$.

Inference is carried out using $10\,000$ iterations of a Markov chain including $1\,500$ burn-in samples.
The parameters of the proposed sampler are set to $\alpha = 0.99$, $\eta = 10^{-5}$ and $\epsilon = 10^{-6}$ for PMALA, and to $p = 0.5$ and $K = 50$ for MTM.
Since the operator $\Delta$ only compares a pixel to its four neighbors and since the indicator prior and likelihood are pixel-wise, the set of pixels can be partitioned into two conditionally independent subsets of pixels.
A two sites Chromatic Gibbs sampling~\cite{gonzalez_parallel_2011} is therefore performed in the MTM kernel to speed up computations.
Note that using the smooth uniform prior as a proposal distribution in MTM is inefficient due to the small size of high probability regions compared to the volume of $\fvalidity$.
The proposal distribution $q$ is based on the spatial prior~\eqref{eq:spatial_prior} instead.
For any pixel, one can show that the conditional spatial prior
is a Gaussian distribution centered on the mean of the set of neighboring pixels $V_n$.
Since maps are assumed to be smooth, the likelihood functions for a pixel $n$ and its neighbors should correspond to similar modes in the parameters' domain.
If the neighbors are not all in the same mode, the mean of the neighbors will in general not fall in a high probability region.
Therefore, for a pixel $n$, the proposal distribution is defined as a Gaussian mixture whose modes are all the means of non-empty subsets $V\in\mathcal{P}(V_n)$ of $V_n$:
%
%
%
\begin{align}
    q (\paramvect{n}  \vert \paramfull_{\setminus n})
    \propto \prod_{d=1}^D 
    \sum_{V \in \mathcal{P}(V_n)} \hspace{-3mm}
    \exp \hspace{-1mm} \left[
        -2 \tau_d \sum_{i \in V} (\theta_{nd} - \theta_{id})^2
    \right] \label{eq:proposal_sum_squares} \\
    \propto \prod_{d=1}^D 
    \sum_{V \in \mathcal{P}(V_n)} \hspace{-3mm}
    \exp \left[
        -2 \tau_d \vert V \vert \left(\theta_{nd} - \frac{1}{\vert V \vert } \sum_{i \in V} \theta_{id} \right)^2
    \right] \label{eq:proposal_means}
    .
\end{align}
%
%
%
%
%

Performance is assessed for the Minimum Mean Squared Error (MMSE) estimate $\widehat{\paramfull}$.
Recall that the inferred parameters $\paramfull$ correspond to normalized logarithms of physical parameters $\Phibf$. Therefore, prediction errors on the $D$ parameter maps $\paramfull_{\cdot d}$ are comparable.
%
%
The quality of the reconstruction is quantified with the Mean Squared Error~(MSE) $\Vert \widehat{\paramfull} - \paramfull^* \Vert_2^2$ and the Reconstruction Signal-to-Noise Ratio~(R-SNR) $20 \log_{10} \Big( \frac{\Vert \paramfull^* \Vert}{\Vert \widehat{\paramfull} - \paramfull^* \Vert}\Big)$.
%

%
%
Fig.~\ref{fig:astro_results} shows the estimations results. 
The MMSE estimate $\widehat{\paramfull}$ (middle) is very close to the ground truth $\paramfull^*$ (left).
The reconstructions are qualitatively very consistent with the underlying physics.
%
The parameter $\paramfull_{\cdot 4}$, corresponding to $\varphi_4=A_V$, is known by astrophysicists to be the most difficult to retrieve as high values lead to saturated line intensities. 
%
Such pixels appear in the top left corner of the ground truth map.

Table~\ref{tab:results_astro} shows the MSE and the R-SNR for each parameter $\paramfull_{\cdot d}$, and the relative size of the credibility intervals with respect to the associated (normalized) validity interval $\cal{C}$.
As expected, the MSE is larger for $\paramfull_{\cdot 4}$ ($\leftrightarrow \varphi_4=A_V$), and the relative size of its credibility intervals are overall the largest, about 16.2\%.
%
The problem is also very ill-posed for all parameters in pixels with very low SNR, where most of the lines are censored, see Fig.~\ref{fig:astro_obs} (right). 
To interpret the results from an astrophysical viewpoint, performances are computed over two subsets of pixels with either less or more than 50\% of censored lines.
As expected, the credibility intervals of the latter are about twice as large as the former.
%
Finally, all the parameters but $A_V$ are well constrained for pixels with less than 50\% of censored lines.
The inference remains challenging since the posterior contains many local modes with high $g$ values, but the proposal distribution $q$ permits the Markov chain to successfully reach the mode of interest.
The relative quadratic error results in an R-SNR between 15.5 dB and 23.4 dB.
Credibility intervals at 95\% level remain small, ranging from $5.7\%$ to $9.3\%$ of the admissible interval $\cal{C}$.

Combining all the difficulties addressed in this article, this astrophysical inverse problem illustrate the good performances of the proposed approach in a challenging scenario.
The proposed likelihood approximation enabled handling the censorship and mixture of noises present in the observation model.
Dealing with a multimodal posterior distribution, the MTM kernel allows the different modes to be visited, while the PMALA kernel permits to explore them efficiently. The proposed sampler provides high quality estimates and informative credibility intervals.

\begin{table}[!t]
    \begin{center}
    \caption{Reconstruction metrics and relative size of credible intervals for the astrophysics experiment.
    The R-SNR is not defined for $\paramfull_{\cdot 1}$, as its ground truth is 0 everywhere.
    }
    \label{tab:results_astro}
    \begin{tabular}{|ccc|ccc|}
    \hline
    & \multicolumn{2}{c|}{MMSE}
    & \multicolumn{3}{c|}{Mean 95\% credibility intervals size} \\
    \hline
    & \multirow{2}{*}{MSE} & R-SNR
    & \multicolumn{2}{c}{censorship}
    & \multirow{2}{*}{overall} \\
    & & (dB) & $\leq$ 50\% & $>$ 50\% & \\
    \hline
    $\paramfull_{\cdot 1}$ & $0.017$ & -- & $6.1$ \% & $11.9$ \% & $6.8$ \% \\
    $\paramfull_{\cdot 2}$ & $0.019$ & $16.8$ & $9.3$ \% & $20.6$ \% & $9.9$ \%\\
    $\paramfull_{\cdot 3}$ & $0.009$ & $23.4$ & $5.7$ \% & $19.8$ \% & $6.5$ \% \\
    $\paramfull_{\cdot 4}$ & $0.034$ & $15.5$ & $16.3$ \% & $14.5$ \% & $16.2$ \% \\
    \hline
    \end{tabular}
    \end{center}
\end{table}

\section{Conclusion}
\label{section:conclu}

This work addresses a family of inverse problems that combine several difficulties: a non-linear black-box forward model, potentially non-injective, that covers multiple decades; observations damaged by both censorship and a mixture of additive and multiplicative noises.
The likelihood is intractable and leads to a potentially multimodal posterior distribution.
%
%
An approximation of the likelihood was proposed, based on a model reduction and an approximate parametric noise mixture model with controlled error.
To efficiently sample from the resulting multimodal posterior, an original MCMC algorithm combining two kernels was proposed.
The Gibbs-like MTM kernel permits jumps between modes, while the PMALA kernel efficiently explores the local geometry of each mode.
The proposed sampler was shown to be competitive with state-of-the-art multimodal sampling methods on a Gaussian mixture model and a sensor localization problem.
Motivated by astronomical observation, a more realistic application to a challenging inverse problem has shown the interest and the good performances of the proposed approach. Estimation errors remain small and uncertainties are quantified. 
Future work includes applications to real astrophysical data such as the IRAM's Orion-B cloud observations~\cite{pety_anatomy_2016} or the James Webb Spatial Telescope observations.




\section*{Acknowledgments}

This work was partly supported
by the CNRS 80Prime project OrionStat,
by the ANR project ``Chaire IA Sherlock'' ANR-20-CHIA-0031-01 held by P. Chainais,
by the Programme National “Physique et Chimie du Milieu Interstellaire” (PCMI) of CNRS/INSU with INC/INP, co-funded by CEA and CNES,
and by the national support within the {\em programme d'investissements d'avenir} ANR-16-IDEX-0004 ULNE and Région HDF.
We also thank all the ORION-B Consortium (\url{https://www.iram.fr/~pety/ORION-B/team.html}).

\appendices

\section{Choice of likelihood approximation parameters $\aell$}
\label{section:llh_param_optim}

For each channel $\ell$, the parameter $\aell = (\lambdaLowerLimit, \lambdaUpperLimit)$ locates the frontiers between low, intermediate and high values regimes of $\approxPell$ in the definition of $\lambda$~\eqref{eq:def_lambda}.
It has a critical influence on the approximation quality.
It should be adjusted to $\approxPell$, $\sigma_a$ and $\sigma_m$.
%
For simplicity, in this subsection, likelihood functions are conditioned with respect to $z = \approxPell(\paramvect{}) \in \R$ instead of $\paramvect{} \in \R^D$.
The true likelihood is not explicit, but the model~\eqref{eq:model} can be easily sampled from, and the approximation~\eqref{eq:likelihood} is known.

The parameter $\aell$ is set to obtain an approximation as close as possible to the true likelihood, with respect to some divergence criterion.
%
%
The Kullback-Leibler (KL) divergence would be a natural choice.
However, due to the number of decades spanned, the standard deviation of KL estimators is in practice larger than the quantity of interest~\cite{kraskov_estimating_2004}, which prevents from performing optimization.
%
%
%
The Kolmogorov-Smirnov (KS) distance is not affected by this property: for a given $z$, it only requires ordered samples $(y^{(i)} )_{i=1}^M$.
%
It reads
\begin{align}
    \widehat{D}_{\text{KS}} (z, \aell)
    =
    \sup_{y \in \R} \left\vert
        \widehat{F}_{M} 
            (y \vert z)
        - \widetilde{F}
            (y \vert z, \aell)
    \right\vert
    ,
\end{align}
where $\widehat{F}_M(\cdot \vert z)$ is the empirical
cdf
of the true likelihood $\pi(\cdot \vert z)$ estimated from $M$ samples $y^{(i)}$, and $\widetilde{F}(\cdot \vert z, \aell)$ is the cdf of the proposed approximation~\eqref{eq:likelihood}.
Assuming that $\paramvect{}$ follows a uniform distribution on $\fvalidity$ yields a distribution on $z$ with pdf $\pi(z)$ which can be estimated by kernel density estimation (KDE).
The function to minimize is
\begin{align}
    \varphi(\aell)
    = \mathbb{E}_{z} \left[
        \widehat{D}_{\text{KS}} \left( z, \aell \right)
    \right]
    = \int \widehat{D}_{\text{KS}} (z, \aell ) \pi(z) dz
    .
\end{align}
An estimator $\widehat{\varphi}$ can be obtained using numerical integration on $z$ over $S$ bins.
%
%
The higher $M$ and $S$, the better the estimation accuracy.
Minimizing $\widehat{\varphi}$ can be performed using a grid search, which is quite computationally intensive.
A cheaper alternative is to use a Bayesian Optimization (BO) procedure~\cite{shahriari_taking_2016}.
This optimization was applied for each channel in the astrophysical application described in Section~\ref{subsec:astro_ex}.
Both grid search and BO approaches were used.
The KDE of $\pi(z)$ was performed from $810\,000$ samples.
The BO procedure was run with $S = 100$ and $M = 250\,000$ using~\cite{code_bo} with default parameters.
Fig.~\ref{fig:optimal_approximation} shows the results for one channel.
The proposed approximation with adjusted $\aell$ is closer to the true likelihood than a purely additive Gaussian approximation, i.e., $\lambdaLowerLimit > \max_j z^{(j)}$, or a purely multiplicative lognormal approximations, i.e., $\lambdaUpperLimit < \min_j z^{(j)}$.

\begin{figure}[!t]
    \subfloat{
        \includegraphics[width=0.24\textwidth, height=2.85cm]{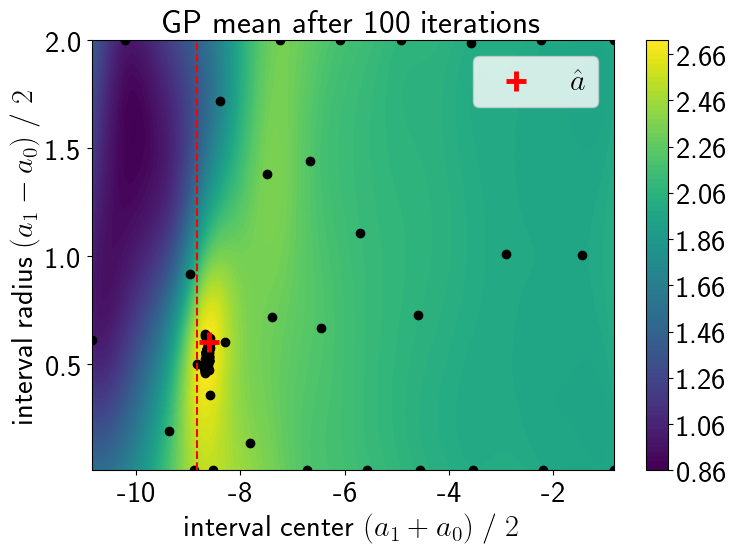}
    }
    \subfloat{
        \includegraphics[width=0.24\textwidth, height=2.85cm]{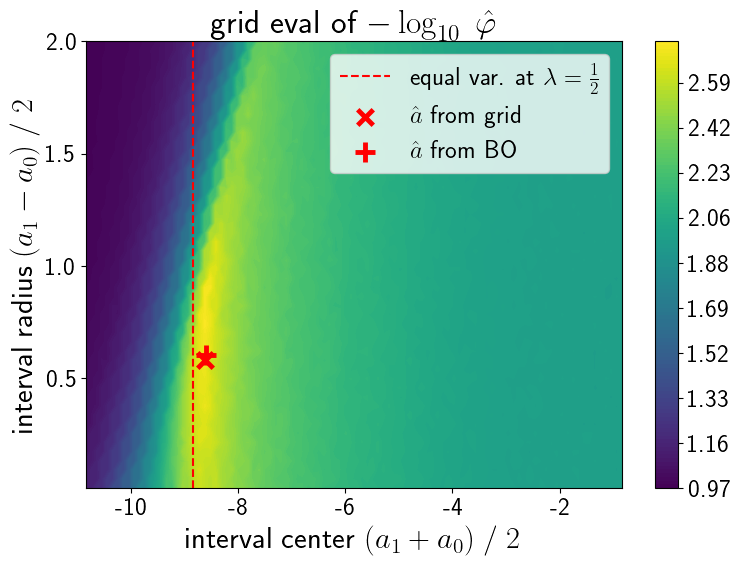}
    } \\
    \subfloat{
        \includegraphics[width=0.24\textwidth, height=2.85cm]{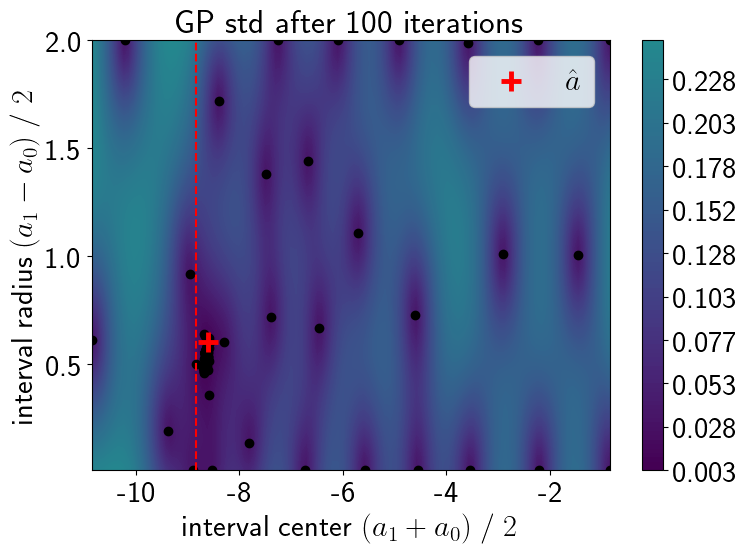}
    }
    \subfloat{
        \includegraphics[width=0.24\textwidth, height=2.85cm]{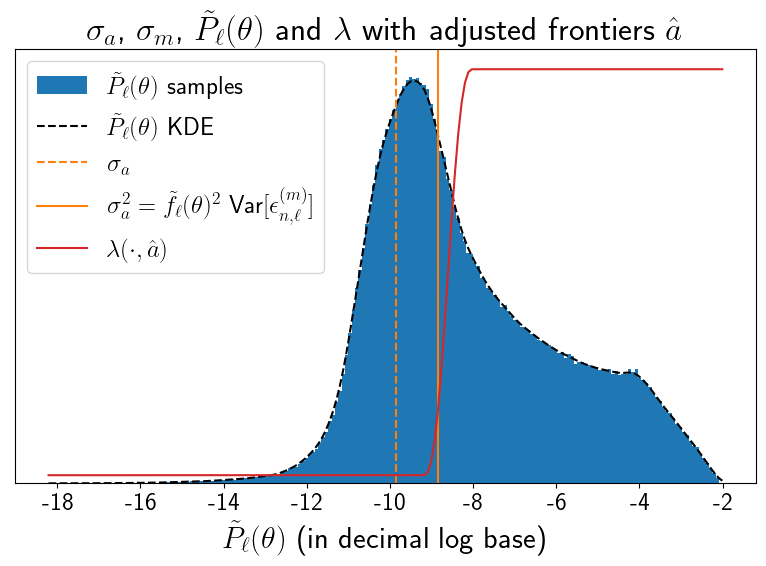}
    }
    \vspace{-1.5mm}
    \caption{
        Maximization of $- \log_{10} \widehat{\varphi}$ 
        using both Bayesian Optimization (BO) and grid search for one channel of the astrophysical case detailed in~\ref{subsec:astro_ex}.
        In BO, a Gaussian Process (GP) replaces the function to optimize (left column).
        The red dashed vertical bar represents the value of $\frac{a_0 + a_1}{2}$ for which the additive and multiplicative noises have equal variances, i.e. $\sigma_a^2 = \tilde{f}_\ell (\paramvect{})^2 \text{Var}[ \epsilon^{(m)}_{n,\ell} ]$, at $\lambda = \frac{1}{2}$.
        For clarity, all scales are displayed in $\log_{10}$ scale, while computations are done in $\log$ scale.
    }
    \label{fig:optimal_approximation}
\end{figure}

\section{Sampling from smoothed indicator distribution}
\label{section:sampling_smmoth_indicator}

This section describes the algorithm to draw samples from the real-valued probability distribution with density $\pi( \theta ) \propto \exp\left( - \delta \, \tilde{\iota}_{[l, u]}(\theta) \right)$, with $l < u$ and $\tilde{\iota}_{[l, u]}$ introduced in~\eqref{eq:_neg_log_smooth_indicator_prior}.
To this aim, consider the generalized normal distribution $G\mathcal{N}(0, 1/\delta^4, 4)$ of probability density function~\cite{nadarajah_generalized_2005}
\begin{align}
    p_{G\mathcal{N}}(\theta)
    =
    \frac{2 \delta^{\frac{1}{4}}}{ \Gamma(1/4)} \exp \left(
        - \delta \, \theta^4
    \right).
\end{align}
Note that $\pi(\theta)$ is a continuous extension of a uniform distribution and of this generalized normal distribution at $0$.
\begin{align}
    \label{eq:smooth_indicator}
    \pi( \theta )
    \propto
    \begin{cases}
        p_{G\mathcal{N}}\left(\theta - l\right) & \text{ if } \theta < l, \\
        p_{G\mathcal{N}}\left(0\right) & \text{ if } \theta \in [l,u], \\
        p_{G\mathcal{N}}\left(u - \theta\right) & \text{ if } \theta > u.
    \end{cases}
\end{align}
The normalizing constant of $\pi(\theta)$ is $1 + p_{G\mathcal{N}}(0) (u - l)$.
The weight of the uniform section in the combination is therefore
\begin{align}\label{eq:weight_unif}
    w_{\text{Unif}}
    =
    \frac{1}{1 + \frac{ \Gamma(1 / 4)}{2} \frac{1}{\delta^{\frac{1}{4}} (u - l)}} \;.
\end{align}
%
Algorithm~\ref{alg:sampleSmoothIndicator} summarizes the procedure to sample from $\pi(\theta)$.

\begin{algorithm}[!t]
    \caption{Sampling from smooth distribution~\eqref{eq:smooth_indicator}}
    \label{alg:sampleSmoothIndicator}
    %
    \DontPrintSemicolon
    \KwIn{scale factor $\delta$, bounds $l,u \in \R$ such that $u > l$}
    \KwOut{sample $\theta$}
    \vspace{1mm}
    $w_{\text{Unif}}$
    \tcp*{using \eqref{eq:weight_unif}}
    $z \sim \mathcal{B}(w_{\text{Unif}})$ \;
    \textbf{if} $z = 1$ \textbf{then} $\theta \sim \text{Unif}([l, u])$ \;
    \Else{
        $\theta \sim G\mathcal{N}(0, 1/\delta^4, 4)$
        \tcp*{using~\cite{nardon_simulation_2009}}
        \textbf{if} $\theta < 0$
        \textbf{then} $\theta = \theta + l$
 \textbf{else} $\theta = \theta + u$ \;
    }
\end{algorithm}


\bibliographystyle{IEEEtran}
\bibliography{IEEEabrv,biblio,biblio_code}

\begin{thebibliography}{10}
\providecommand{\url}[1]{#1}
\csname url@samestyle\endcsname
\providecommand{\newblock}{\relax}
\providecommand{\bibinfo}[2]{#2}
\providecommand{\BIBentrySTDinterwordspacing}{\spaceskip=0pt\relax}
\providecommand{\BIBentryALTinterwordstretchfactor}{4}
\providecommand{\BIBentryALTinterwordspacing}{\spaceskip=\fontdimen2\font plus
\BIBentryALTinterwordstretchfactor\fontdimen3\font minus \fontdimen4\font\relax}
\providecommand{\BIBforeignlanguage}[2]{{%
\expandafter\ifx\csname l@#1\endcsname\relax
\typeout{** WARNING: IEEEtran.bst: No hyphenation pattern has been}%
\typeout{** loaded for the language `#1'. Using the pattern for}%
\typeout{** the default language instead.}%
\else
\language=\csname l@#1\endcsname
\fi
#2}}
\providecommand{\BIBdecl}{\relax}
\BIBdecl

\bibitem{li_preconditioned_2016}
C.~Li, C.~Chen, D.~Carlson \emph{et~al.}, ``\BIBforeignlanguage{en}{Preconditioned {Stochastic} {Gradient} {Langevin} {Dynamics} for {Deep} {Neural} {Networks}},'' \emph{\BIBforeignlanguage{en}{AAAI}}, vol.~30, no.~1, Feb. 2016.

\bibitem{liu_multiple-try_2021}
J.~S. Liu, F.~Liang, and W.~H. Wong, ``\BIBforeignlanguage{en}{The {Multiple}-{Try} {Method} and {Local} {Optimization} in {Metropolis} {Sampling}},'' p.~15, 2021.

\bibitem{walker_parameter_2010}
D.~M. Walker, D.~Allingham, H.~W.~J. Lee \emph{et~al.}, ``\BIBforeignlanguage{en}{Parameter inference in small world network disease models with approximate {Bayesian} {Computational} methods},'' \emph{\BIBforeignlanguage{en}{Physica A: Statistical Mechanics and its Applications}}, vol. 389, no.~3, pp. 540--548, Feb. 2010.

\bibitem{wu_constraining_2018}
R.~Wu, E.~Bron, T.~Onaka \emph{et~al.}, ``\BIBforeignlanguage{en}{Constraining physical conditions for the {PDR} of {Trumpler} 14 in the {Carina} {Nebula}},'' \emph{\BIBforeignlanguage{en}{A\&A}}, vol. 618, p. A53, Oct. 2018.

\bibitem{joblin_structure_2018}
C.~Joblin, E.~Bron, C.~Pinto \emph{et~al.}, ``\BIBforeignlanguage{en}{Structure of photodissociation fronts in star-forming regions revealed by \textit{{Herschel}} observations of high-{J} {CO} emission lines},'' \emph{\BIBforeignlanguage{en}{A\&A}}, vol. 615, p. A129, Jul. 2018.

\bibitem{le_petit_model_2006}
F.~Le~Petit, C.~Nehme, J.~Le~Bourlot \emph{et~al.}, ``\BIBforeignlanguage{en}{A {Model} for {Atomic} and {Molecular} {Interstellar} {Gas}: {The} {Meudon} {PDR} {Code}},'' \emph{\BIBforeignlanguage{en}{ASTROPHYS J SUPPL S}}, vol. 164, no.~2, pp. 506--529, Jun. 2006.

\bibitem{krissian_oriented_2007}
K.~Krissian, C.-F. Westin, R.~Kikinis \emph{et~al.}, ``Oriented {Speckle} {Reducing} {Anisotropic} {Diffusion},'' \emph{IEEE Transactions on Image Processing}, vol.~16, no.~5, pp. 1412--1424, May 2007.

\bibitem{durand_multiplicative_2010}
S.~Durand, J.~Fadili, and M.~Nikolova, ``\BIBforeignlanguage{en}{Multiplicative {Noise} {Removal} {Using} {L1} {Fidelity} on {Frame} {Coefficients}},'' \emph{\BIBforeignlanguage{en}{J Math Imaging Vis}}, vol.~36, no.~3, pp. 201--226, Mar. 2010.

\bibitem{robert_monte_2004}
C.~P. Robert and G.~Casella, \emph{\BIBforeignlanguage{en}{Monte {Carlo} {Statistical} {Methods}}}, ser. Springer {Texts} in {Statistics}.\hskip 1em plus 0.5em minus 0.4em\relax New York, NY: Springer New York, 2004.

\bibitem{pereyra_survey_2016}
M.~Pereyra, P.~Schniter, E.~Chouzenoux \emph{et~al.}, ``A {Survey} of {Stochastic} {Simulation} and {Optimization} {Methods} in {Signal} {Processing},'' \emph{IEEE Journal of Selected Topics in Signal Processing}, vol.~10, no.~2, pp. 224--241, Mar. 2016.

\bibitem{luengo_survey_2020}
D.~Luengo, L.~Martino, M.~Bugallo \emph{et~al.}, ``A survey of {Monte} {Carlo} methods for parameter estimation,'' \emph{EURASIP Journal on Advances in Signal Processing}, vol. 2020, no.~1, p.~25, May 2020.

\bibitem{roberts_langevin_2002}
G.~O. Roberts and O.~Stramer, ``\BIBforeignlanguage{en}{Langevin {Diffusions} and {Metropolis}-{Hastings} {Algorithms}},'' \emph{\BIBforeignlanguage{en}{Methodology and Computing in Applied Probability}}, vol.~4, no.~4, pp. 337--357, Dec. 2002.

\bibitem{brooks_mcmc_2011}
R.~Neal, ``\BIBforeignlanguage{en}{{MCMC} {Using} {Hamiltonian} {Dynamics}},'' in \emph{\BIBforeignlanguage{en}{Handbook of {Markov} {Chain} {Monte} {Carlo}}}, S.~Brooks, A.~Gelman, G.~Jones \emph{et~al.}, Eds.\hskip 1em plus 0.5em minus 0.4em\relax Chapman and Hall/CRC, May 2011, vol. 20116022.

\bibitem{girolami_riemann_2011}
M.~Girolami and B.~Calderhead, ``\BIBforeignlanguage{en}{Riemann manifold {Langevin} and {Hamiltonian} {Monte} {Carlo} methods},'' \emph{\BIBforeignlanguage{en}{Journal of the Royal Statistical Society: Series B (Statistical Methodology)}}, vol.~73, no.~2, pp. 123--214, 2011.

\bibitem{xifara_langevin_2014}
T.~Xifara, C.~Sherlock, S.~Livingstone \emph{et~al.}, ``\BIBforeignlanguage{en}{Langevin diffusions and the {Metropolis}-adjusted {Langevin} algorithm},'' \emph{\BIBforeignlanguage{en}{Statistics \& Probability Letters}}, vol.~91, pp. 14--19, Aug. 2014.

\bibitem{rmsprop}
T.~Tieleman and G.~Hinton, ``Lecture 6.5-rmsprop: Divide the gradient by a running average of its recent magnitude,'' pp. 26--31, 2012.

\bibitem{ihler_nonparametric_2005}
A.~Ihler, J.~Fisher, R.~Moses \emph{et~al.}, ``Nonparametric belief propagation for self-localization of sensor networks,'' \emph{IEEE Journal on Selected Areas in Communications}, vol.~23, no.~4, pp. 809--819, Apr. 2005.

\bibitem{palud_mixture_2022}
P.~Palud, P.~Chainais, F.~L. Petit \emph{et~al.}, ``Mixture of noises and sampling of non-log-concave posterior distributions,'' in \emph{2022 30th {European} {Signal} {Processing} {Conference} ({EUSIPCO})}, Aug. 2022, pp. 2031--2035.

\bibitem{beaumont_approximate_2002}
M.~A. Beaumont, W.~Zhang, and D.~J. Balding, ``\BIBforeignlanguage{en}{Approximate {Bayesian} {Computation} in {Population} {Genetics}},'' \emph{\BIBforeignlanguage{en}{Genetics}}, vol. 162, no.~4, pp. 2025--2035, Dec. 2002.

\bibitem{peterson_zonal_2017}
J.~L. Peterson, K.~D. Humbird, J.~E. Field \emph{et~al.}, ``\BIBforeignlanguage{en}{Zonal flow generation in inertial confinement fusion implosions},'' \emph{\BIBforeignlanguage{en}{Phys. Plasmas}}, vol.~24, no.~3, p. 032702, Mar. 2017.

\bibitem{kwan_cosmic_2015}
J.~Kwan, K.~Heitmann, S.~Habib \emph{et~al.}, ``Cosmic {Emulation}: {Fast} {Predictions} for the {Galaxy} {Power} {Spectrum},'' \emph{The Astrophysical Journal}, vol. 810, p.~35, Sep. 2015.

\bibitem{kasim_building_2021}
M.~F. Kasim, D.~Watson-Parris, L.~Deaconu \emph{et~al.}, ``\BIBforeignlanguage{en}{Building high accuracy emulators for scientific simulations with deep neural architecture search},'' \emph{\BIBforeignlanguage{en}{Mach. Learn.: Sci. Technol.}}, vol.~3, no.~1, p. 015013, Dec. 2021.

\bibitem{bobin_non-linear_2021}
J.~Bobin, R.~C. Gertosio, C.~Bobin \emph{et~al.}, ``Non-linear interpolation learning for example-based inverse problem regularization,'' Jun. 2021.

\bibitem{huang_convex_2013}
Y.~Huang, M.~Ng, and T.~Zeng, ``\BIBforeignlanguage{en}{The {Convex} {Relaxation} {Method} on {Deconvolution} {Model} {withMultiplicative} {Noise}},'' \emph{\BIBforeignlanguage{en}{Communications in Computational Physics}}, vol.~13, no.~4, pp. 1066--1092, Apr. 2013.

\bibitem{nicholson_additive_2020}
R.~Nicholson and J.~P. Kaipio, ``\BIBforeignlanguage{en}{An {Additive} {Approximation} to {Multiplicative} {Noise}},'' \emph{\BIBforeignlanguage{en}{J Math Imaging Vis}}, vol.~62, no.~9, pp. 1227--1237, Nov. 2020.

\bibitem{nocedal_numerical_2006}
J.~Nocedal and S.~J. Wright, \emph{\BIBforeignlanguage{en}{Numerical optimization}}, 2nd~ed., ser. Springer series in operations research.\hskip 1em plus 0.5em minus 0.4em\relax New York: Springer, 2006.

\bibitem{metropolis_equation_1953}
N.~Metropolis, A.~W. Rosenbluth, M.~N. Rosenbluth \emph{et~al.}, ``Equation of {State} {Calculations} by {Fast} {Computing} {Machines},'' \emph{J. Chem. Phys.}, vol.~21, no.~6, pp. 1087--1092, Jun. 1953.

\bibitem{hastings_monte_1970}
W.~K. Hastings, ``Monte {Carlo} {Sampling} {Methods} {Using} {Markov} {Chains} and {Their} {Applications},'' \emph{Biometrika}, vol.~57, no.~1, pp. 97--109, 1970.

\bibitem{kou_equi-energy_2006}
S.~C. Kou, Q.~Zhou, and W.~H. Wong, ``Equi-energy sampler with applications in statistical inference and statistical mechanics,'' \emph{The Annals of Statistics}, vol.~34, no.~4, pp. 1581--1619, Aug. 2006.

\bibitem{miasojedow_adaptive_2013}
B.~Miasojedow, E.~Moulines, and M.~Vihola, ``An {Adaptive} {Parallel} {Tempering} {Algorithm},'' \emph{Journal of Computational and Graphical Statistics}, vol.~22, no.~3, pp. 649--664, 2013.

\bibitem{andricioaei_smart_2001}
I.~Andricioaei, J.~E. Straub, and A.~F. Voter, ``Smart {Darting} {Monte} {Carlo},'' \emph{J. Chem. Phys.}, vol. 114, no.~16, pp. 6994--7000, Apr. 2001.

\bibitem{pompe_framework_2020}
E.~Pompe, C.~Holmes, and K.~Latuszynski, ``A framework for adaptive {MCMC} targeting multimodal distributions,'' \emph{Annals of Statistics}, vol.~48, pp. 2930--2952, Oct. 2020.

\bibitem{ahn_distributed_2013}
S.~Ahn, Y.~Chen, and M.~Welling, ``\BIBforeignlanguage{en}{Distributed and {Adaptive} {Darting} {Monte} {Carlo} through {Regenerations}},'' in \emph{\BIBforeignlanguage{en}{Proceedings of the {Sixteenth} {International} {Conference} on {Artificial} {Intelligence} and {Statistics}}}.\hskip 1em plus 0.5em minus 0.4em\relax PMLR, Apr. 2013, pp. 108--116.

\bibitem{lan_wormhole_2014}
S.~Lan, J.~Streets, and B.~Shahbaba, ``Wormhole {Hamiltonian} {Monte} {Carlo},'' in \emph{Proceedings of the {Twenty}-{Eighth} {AAAI} {Conference} on {Artificial} {Intelligence}}, ser. {AAAI}'14.\hskip 1em plus 0.5em minus 0.4em\relax Québec City, Québec, Canada: AAAI Press, Jul. 2014, pp. 1953--1959.

\bibitem{gilks_adaptive_1998}
W.~R. Gilks, G.~O. Roberts, and S.~K. Sahu, ``Adaptive {Markov} {Chain} {Monte} {Carlo} through {Regeneration},'' \emph{Journal of the American Statistical Association}, vol.~93, no. 443, pp. 1045--1054, 1998.

\bibitem{martino_review_2018}
L.~Martino, ``\BIBforeignlanguage{en}{A review of multiple try {MCMC} algorithms for signal processing},'' \emph{\BIBforeignlanguage{en}{Digital Signal Processing}}, vol.~75, pp. 134--152, Apr. 2018.

\bibitem{martino_flexibility_2013}
L.~Martino and J.~Read, ``\BIBforeignlanguage{en}{On the flexibility of the design of multiple try {Metropolis} schemes},'' \emph{\BIBforeignlanguage{en}{Comput Stat}}, vol.~28, no.~6, pp. 2797--2823, Dec. 2013.

\bibitem{gonzalez_parallel_2011}
J.~Gonzalez, Y.~Low, A.~Gretton \emph{et~al.}, ``\BIBforeignlanguage{en}{Parallel {Gibbs} {Sampling}: {From} {Colored} {Fields} to {Thin} {Junction} {Trees}},'' in \emph{\BIBforeignlanguage{en}{Proceedings of the {Fourteenth} {International} {Conference} on {Artificial} {Intelligence} and {Statistics}}}.\hskip 1em plus 0.5em minus 0.4em\relax JMLR Workshop and Conference Proceedings, Jun. 2011, pp. 324--332.

\bibitem{jones_convergence_2014}
G.~L. Jones, G.~O. Roberts, and J.~S. Rosenthal, ``Convergence of {Conditional} {Metropolis}-{Hastings} {Samplers},'' \emph{Advances in Applied Probability}, vol.~46, no.~2, pp. 422--445, 2014.

\bibitem{roberts_geometric_1996}
G.~O. Roberts and R.~L. Tweedie, ``Geometric {Convergence} and {Central} {Limit} {Theorems} for {Multidimensional} {Hastings} and {Metropolis} {Algorithms},'' \emph{Biometrika}, vol.~83, no.~1, pp. 95--110, 1996.

\bibitem{pety_anatomy_2016}
J.~Pety, V.~Guzman, J.~Orkisz \emph{et~al.}, ``The anatomy of the {Orion} {B} {Giant} {Molecular} {Cloud}: {A} local template for studies of nearby galaxies,'' \emph{Astronomy \& Astrophysics}, vol. 599, Nov. 2016.

\bibitem{kraskov_estimating_2004}
A.~Kraskov, H.~Stögbauer, and P.~Grassberger, ``Estimating mutual information,'' \emph{Phys. Rev. E}, vol.~69, no.~6, p. 066138, Jun. 2004.

\bibitem{shahriari_taking_2016}
B.~Shahriari, K.~Swersky, Z.~Wang \emph{et~al.}, ``Taking the {Human} {Out} of the {Loop}: {A} {Review} of {Bayesian} {Optimization},'' \emph{Proceedings of the IEEE}, vol. 104, no.~1, pp. 148--175, Jan. 2016.

\bibitem{code_bo}
F.~Nogueira, ``{Bayesian Optimization}: Open source constrained global optimization tool for {Python},'' 2014--.

\bibitem{nadarajah_generalized_2005}
S.~Nadarajah, ``\BIBforeignlanguage{en}{A generalized normal distribution},'' \emph{\BIBforeignlanguage{en}{Journal of Applied Statistics}}, vol.~32, no.~7, pp. 685--694, Sep. 2005.

\bibitem{nardon_simulation_2009}
M.~Nardon and P.~Pianca, ``\BIBforeignlanguage{en}{Simulation techniques for generalized {Gaussian} densities},'' \emph{\BIBforeignlanguage{en}{Journal of Statistical Computation and Simulation}}, vol.~79, no.~11, pp. 1317--1329, Nov. 2009.

\end{thebibliography}

\end{document}